\documentclass[format=acmlarge, review=false, screen=true, natbib=false]{acmart}

\usepackage{cite}
\usepackage{amsmath,amssymb,amsfonts}
\usepackage{algorithmic}
\usepackage{caption}
\usepackage[export]{adjustbox}
\usepackage{cite}
\usepackage{breakurl}
\usepackage{graphicx}
\usepackage{textcomp}
\usepackage{multirow}
\usepackage{multicol}
\usepackage{enumitem}
\usepackage[para]{threeparttable}
\usepackage{tablefootnote}
\usepackage{tabularx}
\usepackage{longtable,color,psfrag,dsfont,supertabular,bm,units}
\usepackage{makecell}
\usepackage{slashbox}
\usepackage{subfig}
\usepackage{array}
\usepackage{comment}
\usepackage{booktabs}
\usepackage{tikz}
\usetikzlibrary{calc}
\tikzset{circle mark/.pic={
  \draw [path picture={%
    \fill (path picture bounding box.south west) |-  
      ($(path picture bounding box.south east)!#1/100!(path picture bounding box.north east)$) 
      |- cycle;}] circle [radius=1ex];
}}
\makeatletter
\newcommand*{\rom}[1]{\expandafter\@slowromancap\romannumeral #1@}
\makeatother
\usepackage{xcolor}
\graphicspath{ {Images/} }

\def\BibTeX{{\rm B\kern-.05em{\sc i\kern-.025em b}\kern-.08emT\kern-.1667em\lower.7ex\hbox{E}\kern-.125emX}}
    
\copyrightyear{2020}
\acmYear{2020}
\setcopyright{acmlicensed}
\acmPrice{15.00}
\acmDOI{10.1145/1122445.1122456}
\acmISBN{978-1-4503-9999-9/18/06}

\begin{document}

\title{A Survey on Security and Privacy Issues in Modern Healthcare Systems: Attacks and Defenses}

\author{AKM Iqtidar Newaz}
\email{anewa001@fiu.edu}
\orcid{0002-8814-3975}
\author{Amit Kumar Sikder}
\email{asikd003@fiu.edu}
\author{Mohammad Ashiqur Rahman}
\email{marahman@fiu.edu}

\author{A. Selcuk Uluagac}
\email{suluagac@fiu.edu}
\affiliation{%
  \institution{Florida International University,USA}
  \streetaddress{10555 West Flagler St. EC 3900}
  \city{Miami}
  \state{Florida}
  \postcode{33174}
}

\begin{abstract}

The recent advancements in computing systems and wireless communications have made healthcare systems more efficient than before. 
Modern healthcare devices can monitor and manage different health conditions of the patients automatically without any manual intervention from the medical professionals. Additionally, the use of implantable medical devices (IMDs), body area networks (BANs), and Internet of Things (IoT) technologies in healthcare systems improve the overall patient monitoring and treatment process. However, these systems are complex in software and hardware, and optimizing between security, privacy, and treatment is crucial for healthcare systems as any security or privacy violation can lead to severe effects on patients' treatments and overall health conditions. 
Indeed, the healthcare domain is increasingly facing security challenges and threats due to numerous design flaws and the lack of proper security measures in healthcare devices and applications. In this paper, we explore various security and privacy threats to healthcare systems and discuss the consequences of these threats. We present a detailed survey of different potential 
attacks and discuss their impacts. Furthermore, we review the existing security measures proposed for healthcare systems and discuss their limitations. Finally, we conclude the paper with future research directions toward securing healthcare systems against common vulnerabilities.

\end{abstract}



\maketitle

\section{Introduction}

In recent years, the healthcare domain has experienced 
a myriad of advancements in terms of new technologies and treatment methods. 
Modern healthcare systems 
have changed the lives of patients and medical professionals in many respects. Nowadays, different healthcare applications have been embedded in consumer devices to remotely collect physiological information of a patient and provide automatic treatment. For instance, smartwatches can monitor different body mechanisms like heart rate and electrocardiogram (ECG), smartphones can track the physical activities and sleep apnea, and implanted glucose monitor can automatically control sugar level by injecting insulin to a patient. Moreover, the development of low power wearable biosensors~\cite{pantelopoulos2010survey}, implantable medical devices (IMDs)~\cite{zhang2014trustworthiness}, ultra-low-power body area networks~\cite{kailas2009wireless}, Internet of Things (IoT) technologies~\cite{razaque2019survey}, and numerous lightweight communication protocols~\cite{zhang2009energy} have helped to develop small-scale sense-actuate healthcare devices that can collect and send different physiological values (e.g., blood pressure, heart rate, etc.) from a patient to the medical professionals remotely and instantly to provide better treatments. Indeed, the increasing popularity and diverse utilities of modern healthcare systems have made the healthcare industry grow at a massive rate. The global medical device market is forecasted to grow at a compound annual growth rate of 4.5\% from 2018 to 2023 and is expected to reach \$409.5 billion by 2025~\cite{intro7}.

In the ecosystem of medical devices and applications, modern technologies, such as implantable and wearable medical devices (IWMDs), biosensors, and body area networks (BANs), have certainly enhanced overall healthcare systems for patients and medical professionals. However, these smarter and advanced healthcare systems are ``more'' complex in software and hardware.
Although the adaptation of new technologies in the healthcare domain is at an early stage, several software and hardware defects have already been found, which can lead to possible malicious attacks~\cite{ronquillo2017software, celik2018sensitive, sikder20176thsense, sikder2018survey}. The open-source development platforms and continuous connectivity paves the way for the attackers to exploit the security and privacy in healthcare systems. In recent years, several healthcare security issues have been reported both in the media and the academic community. A story got popular in media that doctors disabled the wireless connectivity 
of a former U.S. Vice President's pacemaker to protect it from being hacked \cite{intro3}. Adding to this story, researchers demonstrated several cyber attacks on commercial IMDs, including attack scenarios of remotely disabling and reprogramming the therapies performed by an implantable cardiac defibrillator (ICD)~\cite{halperin2008pacemakers, li2011hijacking}. Moreover, healthcare/medical devices are remotely exploitable through the communication media \cite{benessa2008protecting, acar2018peek} (e.g., Wi-Fi, Bluetooth Low Energy (BLE), Zigbee, Z-Wave, etc.) and attackers can easily eavesdrop on the communication channel to access the transmitted data~\cite{halperin2008pacemakers}. The lack of standard practice, the need for timely security patches, and the push from the government 
to keep devices and applications secure exacerbate this situation. Because of the catastrophic health consequences, any security issue concerning healthcare systems should be addressed aggressively and proactively. Unfortunately, there is no comprehensive security solution available in the industry and research community to mitigate the emerging cyber attacks on healthcare systems. Researchers have proposed a few countermeasures (e.g., privacy-preserving communication protocols, encrypted databases, etc.) that cannot address the overall attack surface in healthcare systems. Therefore, the security and privacy in healthcare systems require an immediate attention of the security research community, medical device industry, and regulatory bodies~\cite{zhang2014trustworthiness}.

\vspace{3pt}
\noindent\textbf{{Contributions}}: The aim of this survey paper is to provide a comprehensive overview of the security and privacy trends and emerging threats to healthcare systems to facilitate the understanding of the pressing security and privacy challenges. 
The contributions of this work are as follows:
\begin{itemize}[wide=0pt]
        \item First, we provide a detailed overview of a typical healthcare system and discuss its components. 
        \item Second, we explore different security and privacy goals for healthcare systems and discuss potential adversarial models.
        \item Third, we present a detailed taxonomy of the existing 
        attacks in the healthcare domain by analyzing them as reported by the research community and industry. We also discuss the impact of these attacks based on common vulnerability scoring metrics.
        \item Fourth, we summarize the existing solutions that have been proposed 
        to mitigate these attacks and identify the challenges faced by the research community to ensure security and privacy in healthcare systems.
        \item Finally, we articulate several open challenges and future research directions toward solving the security and privacy issues in healthcare systems.
\end{itemize}
    
\vspace{3pt}
\noindent\textbf{{Organization}}: The remainder of this paper is organized as follows: We provide an overview of existing literature surveys on healthcare systems in Section~\ref{related_work}. We present the architecture of a typical healthcare system and discuss its different components in Section~\ref{background}. In the following section, we provide security and privacy goals for healthcare systems. In Section~\ref{threats}, we present a detailed taxonomy of existing security and privacy 
attacks on healthcare systems and summarize the impacts of these attacks based on common vulnerability metrics. In Section~\ref{solutions}, we discuss existing approaches that have been proposed to secure healthcare systems by researchers. Limitations of current security solutions, requirements to form a secured healthcare system and corresponding challenges are discussed in Section~\ref{discussion}. Finally, we conclude the paper in Section~\ref{conclusion}.







\vspace{-0.2cm}
\section{Related Work}\label{related_work}


In recent years, several surveys have been conducted to review existing security and privacy attacks on healthcare systems. However, these works either focus on specific attacks or security solutions for specific devices without considering overall security and privacy issues in healthcare systems. In this section, we summarize these surveys 
and discuss their differences from our work. 
    
\vspace{3pt}
\noindent\textbf{Existing surveys:} 
Existing surveys are mostly focused on security and privacy problems, major vulnerabilities, and solutions related to the privacy and safety issues of IMDs \cite{zhang2014trustworthiness, ellouze2014security, altawy2016security, rathore2017review, camara2015security, kim2015reliability, rushanan2014sok}. 
Among these surveys, Rushanan et al. extensively reviewed security and privacy problems corresponding to telemetry interfaces and software programs, security frameworks, and standard practices that aimed at improving the security of IMDs~\cite{rushanan2014sok}. Other surveys have different focuses. Alemdar et al. evaluated the current research activities and issues that need to be addressed to enhance remote health monitoring for the elderly~\cite{alemdar2010wireless}. David et al. presented a survey on wireless medical sensor networks (WMSNs), 
cryptographic approaches to preserve health data, and the trade-off between security and reliability of WMSNs~\cite{david2016comprehensive}. Sametinger et al. reported the critical issues that were being faced to ensure the safety and security of medical devices and provided an illustrative example~\cite{sametinger2015security}.

\begin{table*}[t!]
\caption{Comparison among our survey and existing surveys.} 
\vspace{-0.5cm}
\begin{tablenotes}
       \footnotesize
      \item \tikz\path pic{circle mark= 0}; = No information provided, \tikz\path pic{circle mark= 50}; = Partial information provided, \tikz\path pic{circle mark= 100}; = Complete information provided
    \end{tablenotes}
\label{table:comparison}
\begin{threeparttable}
\centering
\resizebox{1\textwidth}{!}{
\begin{tabular}{|c|c|c|c|c|c|c|c|c|}
\hline
\multirow{3}{*}{Ref.} & \multirow{3}{*}{\begin{tabular}[c]{@{}c@{}}Components \\of Healthcare \\System\end{tabular}} 
& \multirow{3}{*}{\begin{tabular}[c]{@{}c@{}}Security \\and Privacy \\Goals\end{tabular}} 
& \multicolumn{4}{c|}{\begin{tabular}[c]{@{}c@{}}Attacks on Healthcare Systems\end{tabular}} 
& \multicolumn{2}{c|}{\begin{tabular}[c]{@{}c@{}}Solutions for Existing Attacks\end{tabular}}
 \\ \cline{4-7} \cline{8-9}
 & & & \begin{tabular}[c]{@{}c@{}}Attack\\ Taxonomy\end{tabular} & \begin{tabular}[c]{@{}c@{}}Existing\\ Attacks\end{tabular} & \begin{tabular}[c]{@{}c@{}}Impacted\\ Security\end{tabular} & \begin{tabular}[c]{@{}c@{}}Vulnerability\\ Scoring\end{tabular} & \begin{tabular}[c]{@{}c@{}}Solution\\ Taxonomy\end{tabular} & \begin{tabular}[c]{@{}c@{}}Existing\\ Solutions\end{tabular} \\ \hline
 
 \multicolumn{1}{|c|}{\multirow{1}{*}{\begin{tabular}[c]{@{}c@{}}Rushanan et al. \cite{rushanan2014sok}\end{tabular}}}   & \tikz\path pic{circle mark= 0};        & \tikz\path pic{circle mark= 100}; & \tikz\path pic{circle mark= 0};    & \tikz\path pic{circle mark= 50};   & \tikz\path pic{circle mark= 100};   & \tikz\path pic{circle mark= 0};          & \tikz\path pic{circle mark= 0}; & \tikz\path pic{circle mark= 50};  \\ \hline            
 \multicolumn{1}{|c|}{\multirow{1}{*}{\begin{tabular}[c]{@{}c@{}}Ellouze et al. \cite{ellouze2014security}\end{tabular}}}      & \tikz\path pic{circle mark= 0};        & \tikz\path pic{circle mark= 0}; & \tikz\path pic{circle mark= 0};    & \tikz\path pic{circle mark= 50};   & \tikz\path pic{circle mark= 0};   & \tikz\path pic{circle mark= 0};      & \tikz\path pic{circle mark= 0}; & \tikz\path pic{circle mark= 50}; \\ \hline            
 \multicolumn{1}{|c|}{\multirow{1}{*}{\begin{tabular}[c]{@{}c@{}}Zhang et al. \cite{zhang2014trustworthiness}\end{tabular}}}      & \tikz\path pic{circle mark= 0};        & \tikz\path pic{circle mark= 100}; & \tikz\path pic{circle mark= 50};    & \tikz\path pic{circle mark= 50};   & \tikz\path pic{circle mark= 100};   & \tikz\path pic{circle mark= 0};      & \tikz\path pic{circle mark= 0}; & \tikz\path pic{circle mark= 50}; \\ \hline     
 \multicolumn{1}{|c|}{\multirow{1}{*}{\begin{tabular}[c]{@{}c@{}}Altawy et al. \cite{altawy2016security}\end{tabular}}}      & \tikz\path pic{circle mark= 0};        & \tikz\path pic{circle mark= 50}; & \tikz\path pic{circle mark= 50};    & \tikz\path pic{circle mark= 50};   & \tikz\path pic{circle mark= 100};   & \tikz\path pic{circle mark= 0};      & \tikz\path pic{circle mark= 0}; & \tikz\path pic{circle mark= 50}; \\ \hline     
 \multicolumn{1}{|c|}{\multirow{1}{*}{\begin{tabular}[c]{@{}c@{}}Rathore et al. \cite{rathore2017review}\end{tabular}}}      & \tikz\path pic{circle mark= 0};        & \tikz\path pic{circle mark= 100}; & \tikz\path pic{circle mark= 50};    & \tikz\path pic{circle mark= 50};   & \tikz\path pic{circle mark= 0};   & \tikz\path pic{circle mark= 0};      & \tikz\path pic{circle mark= 0}; & \tikz\path pic{circle mark= 50}; \\ \hline     
 \multicolumn{1}{|c|}{\multirow{1}{*}{\begin{tabular}[c]{@{}c@{}}Camara et al. \cite{camara2015security}\end{tabular}}}      & \tikz\path pic{circle mark= 0};        & \tikz\path pic{circle mark= 100}; & \tikz\path pic{circle mark= 50};    & \tikz\path pic{circle mark= 50};   & \tikz\path pic{circle mark= 0};   & \tikz\path pic{circle mark= 0};      & \tikz\path pic{circle mark= 50}; & \tikz\path pic{circle mark= 50}; \\ \hline     
 \multicolumn{1}{|c|}{\multirow{1}{*}{\begin{tabular}[c]{@{}c@{}}Kim et al. \cite{kim2015reliability}\end{tabular}}}      & \tikz\path pic{circle mark= 0};        & \tikz\path pic{circle mark= 0}; & \tikz\path pic{circle mark= 100};    & \tikz\path pic{circle mark= 50};   & \tikz\path pic{circle mark= 0};   & \tikz\path pic{circle mark= 0};      & \tikz\path pic{circle mark= 0}; & \tikz\path pic{circle mark= 50}; \\ \hline  
 \multicolumn{1}{|c|}{\multirow{1}{*}{\begin{tabular}[c]{@{}c@{}}Alemdar et al. \cite{alemdar2010wireless}\end{tabular}}}      & \tikz\path pic{circle mark= 0};        & \tikz\path pic{circle mark= 100}; & \tikz\path pic{circle mark= 0};    & \tikz\path pic{circle mark= 0};   & \tikz\path pic{circle mark= 0};   & \tikz\path pic{circle mark= 0};      & \tikz\path pic{circle mark= 0}; & \tikz\path pic{circle mark= 50}; \\ \hline  
 \multicolumn{1}{|c|}{\multirow{1}{*}{\begin{tabular}[c]{@{}c@{}}David et al. \cite{david2016comprehensive}\end{tabular}}}      & \tikz\path pic{circle mark= 0};        & \tikz\path pic{circle mark= 0}; & \tikz\path pic{circle mark= 0};    & \tikz\path pic{circle mark= 0};   & \tikz\path pic{circle mark= 0};   & \tikz\path pic{circle mark= 0};      & \tikz\path pic{circle mark= 0}; & \tikz\path pic{circle mark= 50}; \\ \hline  
 \multicolumn{1}{|c|}{\multirow{1}{*}{\begin{tabular}[c]{@{}c@{}}Sametin. et al. \cite{sametinger2015security}\end{tabular}}}      & \tikz\path pic{circle mark= 0};        & \tikz\path pic{circle mark= 0}; & \tikz\path pic{circle mark= 0};    & \tikz\path pic{circle mark= 50};   & \tikz\path pic{circle mark= 0};   & \tikz\path pic{circle mark= 0};      & \tikz\path pic{circle mark= 0}; & \tikz\path pic{circle mark= 50}; \\ \hline 
 \multicolumn{1}{|c|}{\multirow{1}{*}{\begin{tabular}[c]{@{}c@{}}Pantelop. et al. \cite{pantelopoulos2010survey}\end{tabular}}}      & \tikz\path pic{circle mark= 50};        & \tikz\path pic{circle mark= 0}; & \tikz\path pic{circle mark= 0};    & \tikz\path pic{circle mark= 0};   & \tikz\path pic{circle mark= 0};   & \tikz\path pic{circle mark= 0};      & \tikz\path pic{circle mark= 0}; & \tikz\path pic{circle mark= 50}; \\ \hline  
 \multicolumn{1}{|c|}{\multirow{1}{*}{\begin{tabular}[c]{@{}c@{}}Razaque et al. \cite{razaque2019survey}\end{tabular}}}      & \tikz\path pic{circle mark= 50};        & \tikz\path pic{circle mark= 0}; & \tikz\path pic{circle mark= 50};    & \tikz\path pic{circle mark= 50};   & \tikz\path pic{circle mark= 0};   & \tikz\path pic{circle mark= 0};      & \tikz\path pic{circle mark= 50}; & \tikz\path pic{circle mark= 50}; \\ \hline  
 \multicolumn{1}{|c|}{\multirow{1}{*}{\begin{tabular}[c]{@{}c@{}}Habibzadeh et al. \cite{habibzadeh2020toward}\end{tabular}}}      & \tikz\path pic{circle mark= 50};        & \tikz\path pic{circle mark= 50}; & \tikz\path pic{circle mark= 0};    & \tikz\path pic{circle mark= 50};   & \tikz\path pic{circle mark= 0};   & \tikz\path pic{circle mark= 0};      & \tikz\path pic{circle mark= 0}; & \tikz\path pic{circle mark= 50}; \\ \hline  
 \multicolumn{1}{|c|}{\multirow{1}{*}{\begin{tabular}[c]{@{}c@{}}Islam et al. \cite{islam2015internet}\end{tabular}}}      & \tikz\path pic{circle mark= 50};        & \tikz\path pic{circle mark= 100}; & \tikz\path pic{circle mark= 50};    & \tikz\path pic{circle mark= 50};   & \tikz\path pic{circle mark= 0};   & \tikz\path pic{circle mark= 0};      & \tikz\path pic{circle mark= 0}; & \tikz\path pic{circle mark= 50}; \\ \hline  
 \multicolumn{1}{|c|}{\multirow{1}{*}{\begin{tabular}[c]{@{}c@{}}Kruse et al. \cite{kruse2017cybersecurity}\end{tabular}}}      & \tikz\path pic{circle mark= 0};        & \tikz\path pic{circle mark= 0}; & \tikz\path pic{circle mark= 0};    & \tikz\path pic{circle mark= 50};   & \tikz\path pic{circle mark= 0};   & \tikz\path pic{circle mark= 0};      & \tikz\path pic{circle mark= 0}; & \tikz\path pic{circle mark= 50}; \\ \hline  
 \multicolumn{1}{|c|}{\multirow{1}{*}{\begin{tabular}[c]{@{}c@{}}Yaqoob et al. \cite{yaqoob2019security}\end{tabular}}}      & \tikz\path pic{circle mark= 50};        & \tikz\path pic{circle mark= 0}; & \tikz\path pic{circle mark= 0};    & \tikz\path pic{circle mark= 50};   & \tikz\path pic{circle mark= 0};   & \tikz\path pic{circle mark= 0};      & \tikz\path pic{circle mark= 0}; & \tikz\path pic{circle mark= 50}; \\ \hline  
 \multicolumn{1}{|c|}{\multirow{1}{*}{\begin{tabular}[c]{@{}c@{}}Nasiri et al. \cite{nasiri2019security}\end{tabular}}}      & \tikz\path pic{circle mark= 0};        & \tikz\path pic{circle mark= 100}; & \tikz\path pic{circle mark= 0};    & \tikz\path pic{circle mark= 50};   & \tikz\path pic{circle mark= 0};   & \tikz\path pic{circle mark= 0};      & \tikz\path pic{circle mark= 0}; & \tikz\path pic{circle mark= 0}; \\ \hline  
 \multicolumn{1}{|c|}{\multirow{1}{*}{\begin{tabular}[c]{@{}c@{}}Our Survey\end{tabular}}}      & \tikz\path pic{circle mark= 100};        & \tikz\path pic{circle mark= 100}; & \tikz\path pic{circle mark= 100};    & \tikz\path pic{circle mark= 100};   & \tikz\path pic{circle mark= 100};   & \tikz\path pic{circle mark= 100};      & \tikz\path pic{circle mark= 100}; & \tikz\path pic{circle mark= 100}; \\ \hline  

\end{tabular}}
    \end{threeparttable}
    \vspace{-0.5cm}
\end{table*}

A comprehensive review of existing research and development on wearable biosensor systems for health monitoring was presented in~\cite{pantelopoulos2010survey}. Razaque et al. presented a flow of information in the 
healthcare domain  with a particular focus on 
IoT connection \cite{razaque2019survey}. Habibzadeh et al. studied the emerging trends in healthcare applications and discussed potential threats, vulnerabilities, and consequences of cyber attacks in healthcare systems~\cite{habibzadeh2020toward}. Kruse et al. presented a survey on cybersecurity challenges in healthcare systems~\cite{kruse2017cybersecurity}. 
Yaqoob et al. demonstrated possible attack vectors, security vulnerabilities, and applicable attacks were demonstrated for networked medical devices~\cite{yaqoob2019security}. 
Researchers also analyzed IoT security and privacy features from the healthcare perspective. Islam et al. surveyed advances in IoT-based healthcare technologies, network architectures, and industrial trends in IoT-based healthcare solutions~\cite{islam2015internet}. An overview of the features and concepts related to security requirements for IoT in a healthcare system was provided in \cite{nasiri2019security}. 

\vspace{3pt}
\noindent\textbf{Differences from the existing surveys:} 
The main differences between our work and existing surveys can be articulated as follows: (1) While most current surveys are focused on the security and privacy of IMDs and IWMDs 
, our survey focuses on the overall healthcare system,  which covers end-to-end components, including medical devices, sensors, networks/communication, and healthcare providers. (2) We provide a formal architecture on healthcare systems and identify its major components to 
outline security and privacy needs. (3) We categorize the existing security and privacy attacks on healthcare systems and use the common vulnerability scoring system (CVSS) to measure the impact of the attacks. (4) While the existing surveys are hardly focused on the limitations of current security solutions, our work identifies the limitations and discusses them. 
(5) Finally, our survey provides categorical directions for the researchers to explore mitigation measures against common security vulnerabilities in healthcare systems.
We present a comparison among the existing surveys and our survey in Table~\ref{table:comparison}.

\section{Background and definitions}\label{background}

\begin{figure*}[t!]
        \centering
        \includegraphics[width=\textwidth]{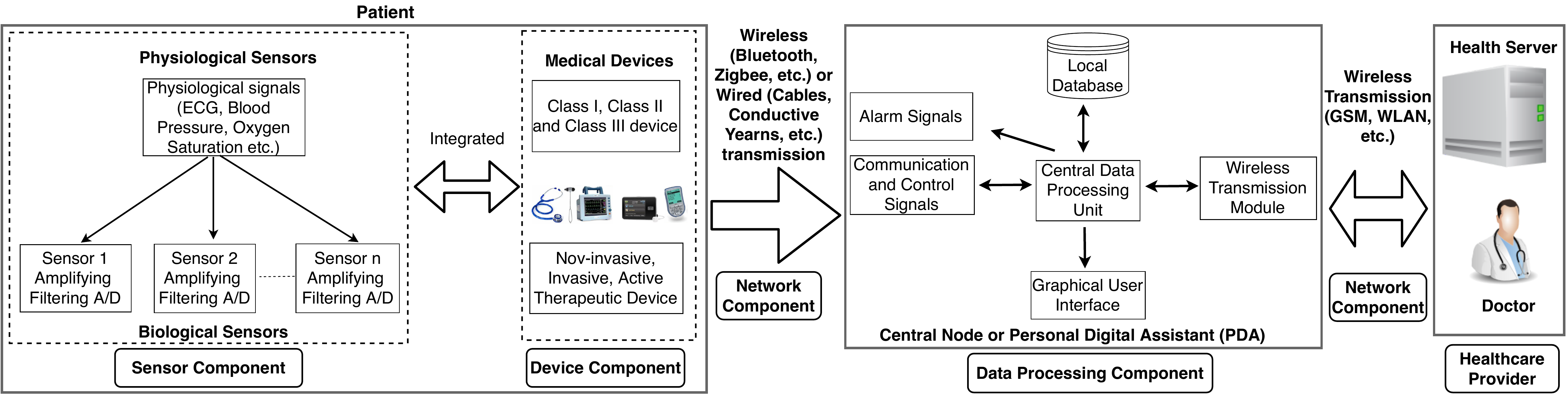}
        \caption{Overview of an example healthcare system.}
        \vspace{-0.3in}
        \label{fig:figure1}
\end{figure*}

In this section, we provide a detailed overview of different components of healthcare systems to understand the significance of security and privacy needs in the healthcare domain. A healthcare system usually comprises one or more medical devices that are equipped with different sensors to collect patients' vitals and takes autonomous decisions to provide enhanced treatments. The overall architecture of a healthcare system is shown in Figure~\ref{fig:figure1}. We identify five major components that are typically important to perform general functionalities of a healthcare system. These five components are medical device, sensor, networking, data processing, and healthcare provider.


\vspace{6pt}
\vspace{-0.2cm}
\noindent\textbf{Medical device:} Any device, instrument, appliance, or apparatus that is intended for one or more medical purposes, such as diagnosis, monitoring, treatment, alleviation, etc., is called a medical device. According to the U.S. food and drug administration (FDA), medical devices range from simple tongue depressors to complex programmable ICD. FDA provides classification standards for medical devices based on the potential risk of causing harm to the patients in case of device malfunction or malicious attacks. Devices with a minimal level of risk and the minimum level of regulatory control, like elastic bandages, dental floss, etc., belong to \textit{Class \rom{1}} medical devices. Pregnancy testing kits, powered wheelchairs, etc. are \textit{Class \rom{2}} devices, which are more complicated and riskier than Class \rom{1} devices and require stringent regulatory controls. \textit{Class \rom{3}} devices (e.g., implantable pacemakers, breast implants, etc.)  possess the highest risk and complexity, and they require highly stringent regulatory controls. In addition, the European Commission provides several other classification standards for medical devices based on \textit{non-invasive}, \textit{invasive}, and \textit{active therapeutic} properties~\cite{background4}:

    

\begin{enumerate} [nosep, wide=0pt, leftmargin=*]
\item \textit{Non-invasive devices:} 
\textit{Non-invasive devices} are intended to use for body-liquid collection in such a way that return flow back to the human body is unlikely (such as urine collection bottles). Also, this type of device only contacts the patient's skin and intends for channeling or storing blood, body-liquids or tissues, liquids, or gases for eventual infusion (e.g., antistatic tubing for anesthesia, syringes for infusion pumps, etc.).

\item \textit{Invasive devices:} These types of devices are introduced into the body, either through a break in the skin or an opening in the body. \textit{Invasive devices} can be further categorized into four groups, namely \textit{transient use, short-term use, long term use, and connected to an active medical device}. Surgically \textit{invasive transient use} medical devices (\textless 60 minutes) are precisely controlled, directly contacted with the central nervous system, and reusable. Surgically  \textit{invasive devices for short-term use} (\textgreater 60 minutes, \textless 30 days) can be directly contacted with the central nervous system to precisely monitor, diagnose, or control the heart central circulatory system. Surgically \textit{invasive long-term use} and \textit{active medical devices} (\textgreater 30 days) can be placed in the mouth, or have direct contact with the heart or central circulatory system to administer medicines.

\item \textit{Active therapeutic devices:} \textit{Active therapeutic devices} are intended to administer or exchange data, whether used alone or in combination with other medical devices, to deliver or remove medicines to or from the body. Examples of such devices include muscle stimulators, dental handpieces, hearing aids, and therapeutic X-ray sources.

\end{enumerate}

\vspace{-0.2cm}    
\vspace{6pt}
\noindent\textbf{Sensor:} In the healthcare domain, sensors are used to monitor and measure the patient's vitals. Different physiological sensors, such as blood sugar sensor, heart rate sensor, etc., are used as a trigger to automate different functionalities (diagnosis, monitoring, etc.) of healthcare systems. We divide sensors into 
the following three categories:
    
\begin{enumerate} [nosep, wide=0pt, leftmargin=*]
        
\item Physiological sensors: 
These sensors measure the physiological signals (e.g., ECG, EMG, etc.) and features to give an overall estimation of the patient's health condition at any given time.
        
\item Biological sensors: 
These sensors integrate the biological elements in a human body with the physio-chemical transducer to produce an electric signal. 
For example, glucose and alcohol sensors are examples of this kind.

\item Environmental sensors: These sensors can sense different environmental parameters to understand any change in the proximity of a patient. For example, an accelerometer and gyroscope in a smartwatch can detect a patient's movement to measure motion and sleep data. 
        
\end{enumerate}

\vspace{6pt}
\noindent\textbf {Networking:} Networking components are concerned with how different medical devices and sensors communicate with each other, 
as well as with other components of a healthcare system. As Figure~\ref{fig:figure1} illustrates, the transmission of measured data in a healthcare system needs to be performed primarily for two different purposes: (1) transferring the physiological signal from the sensors or devices to the system's central node and (2) sending the aggregated measurements from the central node to or from a health server or healthcare professional. Transmission of data for short-range can be handled by wired or wireless channels. However, the wired communication may hinder the patient's mobility and comfortableness \cite{townsend2005recent}. Conductive yearns may be a more favorable approach here to transfer the measurements from sensors integrated on smart-textile clothing~\cite{lam2007recent}. Alternatively, autonomous sensor nodes can follow a primary star topology network to form a 
BAN 
for transmitting data to the central node of the BAN
~\cite{chen2011body}.

The most commonly used wireless communication standards in BANs are IEEE 802.15.1 (Bluetooth) and 802.15.4 (Zigbee), which are a part of the 802.15 working group for wireless personal area network (WPAN). Bluetooth is an industry specification for short-range RF-based connectivity between portable and fixed devices. It is a low-power, low-cost RF standard, operating in the unlicensed 2.4 GHz spectrum~\cite{troster2005agenda}. It uses a frequency hopping technique (FHSS) over 79 channels in the industrial, scientific, and medical (ISM) band to combat interference and supports up to 3 mbps in the enhanced data rate mode with a maximum transmission distance of 100 m. The Zigbee standard also targets low-cost, low data-rate solutions with high battery life. It operates in 16 channels in the 2.4 GHz ISM band (250 kbps, OQPSK modulation), in 10 channels in the 915 MHz band (40 kbps, BPSK modulation) and in one channel in the 868 MHz band (20 kbps, BPSK modulation) \cite{background5}. Alternative technologies for short-range intra-BAN communication include infrared data association (IrDA), ultra-wideband (UWB), and medical implant communication service (MICS). UWB is a low-cost communication protocol for the short-range exchange of data over infrared light. MICS is an ultralow-power, unlicensed, mobile radio service for transmitting low-rate data in support of diagnostic or therapeutic functions associated with medical devices. It uses the 402-405 MHz frequency band, with 300 kHz channels~\cite{yuce2007mics}.

For long-range communication between a healthcare system and a health server or a healthcare provider, there is a wide variety of available wireless technologies (e.g., WLAN, GSM, GPRS, UMTS, WiMAX, LoRa, etc.), which can offer broad coverage and ubiquitous network access. Moreover, future advances in 5G mobile communication systems are expected to guarantee worldwide seamless access to the Internet at much higher data rates, providing the ability to collect data from remote medical devices in real-time. More recently, with the advent of Z-Wave, BLE more devices are expected to be in the market using these low-power communication protocols~\cite{sikder2018iot}.

\vspace{-0.2cm}    
\vspace{6pt}
\noindent\textbf{Data processing:} The data processing component collects data from devices and sensors to produce meaningful information. A central data processing unit is shown in Figure~\ref{fig:figure1}, which communicates with the medical device and sensor components via communication and control modules. 
It has a data processing unit, along with a local database to save initial data about the patient. Its alarm generator informs the patient if there is any anomaly. It uses a wireless transmission module to make a connection with the health server and healthcare provider.

\vspace{-0.2cm}    
\vspace{6pt}
\noindent\textbf{Healthcare provider:} Health servers and healthcare professionals are elements of the healthcare provider component. They communicate with the data processing component through a wireless transmission module. The health server saves healthcare data in the cloud. Healthcare professionals can access this data to treat patients remotely or physically.
    
\vspace{-0.4cm}

\section{Security and privacy needs in existing Healthcare Systems}\label{goals}

\begin{figure*}[t!]
    \centering
    \includegraphics[width=1\textwidth]{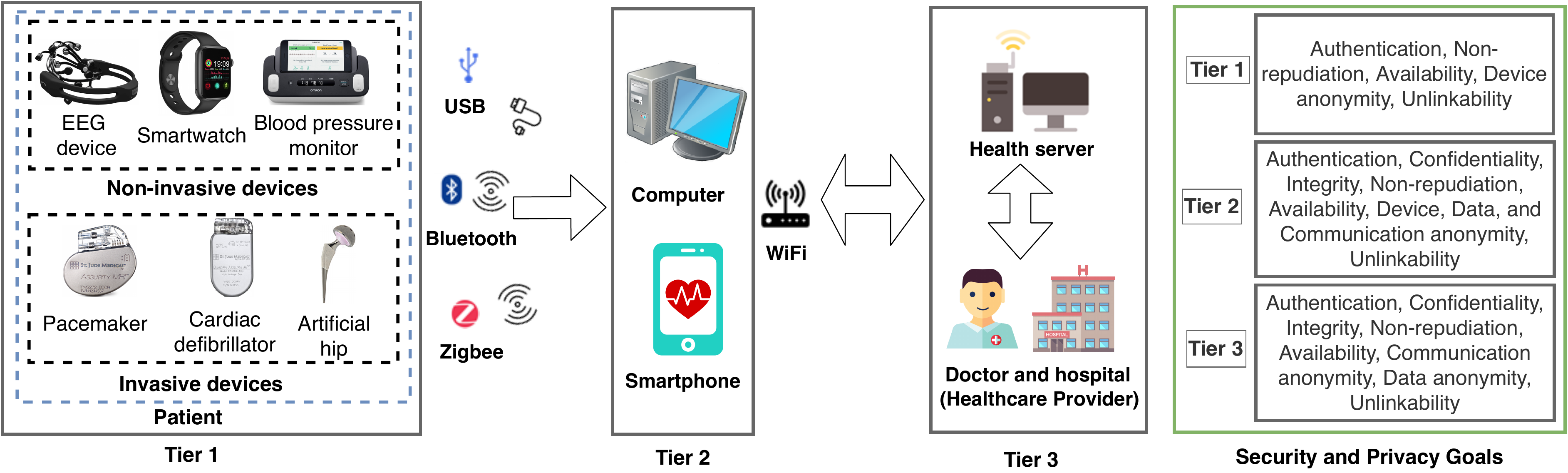}
    \vspace{-0.2in}
    \caption{Security and privacy goals in a healthcare system.}
    \label{fig:figure2}
    \vspace{-0.2in}
\end{figure*}

To discuss the security and privacy issues in healthcare systems, we refer to Figure~\ref{fig:figure2} as the use case scenario, which is a complex multidisciplinary and integrated healthcare system. 
In this section, we first present security and privacy requirements,
and then review the corresponding security and privacy goals.

\vspace{-0.3cm}   
\subsection{Security and Privacy Requirements}
    
Figure~\ref{fig:figure2} presents the general security and privacy goals of a healthcare system. Here, a patient carries several \textit{invasive} and \textit{non-invasive medical devices} that are placed on or around the patient's body to monitor constantly various vital signs of the body (e.g., ECG signal, pulse, blood pressure, etc.) and important environmental parameters (e.g., ambient temperature and humidity). The sensor readings and patient profiles are together called \textit{patient-related} data that is collected and transmitted to other devices like smartphones, computers, etc. These devices can perform further data processing, aggregation, or distributed storage. The patient-related data can also be sent to a central healthcare server for permanent records and to the healthcare professionals and hospital for continuous monitoring of the patient's physical condition. In summary, the overall architecture of a personal healthcare system is divided into three tiers. Tier 1 consists of medical devices, including \textit{invasive} and \textit{non-invasive} devices, and Tier 2 has personal devices like smartphones and computers. Health servers and healthcare professionals form the third tier.

For ensuring security, authentication is required in Tiers 1, 2, and 3. Before transmitting any patient-related data to a personal device or from a personal device to a health server, users must be strictly authorized at each tier. Device information and data should only be accessible by authorized healthcare professionals or the hospital's authority and should not be modifiable by unauthorized users. Confidentiality and integrity should also be ensured at Tiers 2 and 3. As medical devices perform various sensitive operations and handle personal data, this information should be kept confidential in an access log. Medical devices should be reachable all the time as unavailability of the device data may impact the treatment of the patient. Non-repudiation and availability should be maintained in all three tiers. 

Furthermore, to achieve the privacy goals, one needs to maintain device anonymity, which means that only the patients and authorized users should know what medical devices a specific patient is carrying. For transmitting data from Tier 1 to 2 and 2 to 3, patients and doctors' real information should not be disclosed to maintain data anonymity. Communication between the patients and healthcare professionals/hospitals via healthcare system should be untraceable for adversaries to achieve communication anonymity and unlinkability.

\vspace{-0.3cm}   
\subsection{Security and Privacy Goals}
    
We identify the following security goals for healthcare systems based on the previous discussion. To maintain these goals, the following properties should be considered throughout the life cycle of a healthcare system.

\begin{enumerate} [nosep, wide=0pt, leftmargin=*] 
    
\item \textit{Authentication:} Strong authentication is a fundamental component for securing healthcare systems where one needs to consider environment setup, single-factor vs. multi-factor authentication \cite{liu2014maca}, grace periods, and emergency situations. Most of the current 
networked medical devices (e.g., IMDs, IWMDs, etc.) have weak password authentication schemes where password files are stored on the local hard drive \cite{hanna2011take, moses2015lack, zobaed2019clustcrypt}. As a result, an attacker with privileges can delete/modify the files or install new software on the device.
	    
	  \begin{enumerate}[label=(\alph*)]
	   \item Environment considerations: Different healthcare settings have different architectures, and one needs to select the appropriate authentication mechanism accordingly. For example, proximity cards may be suitable and easily accessible for regular patient interaction, but not for authenticating operating rooms.
	        
	   \item Single vs. multi-factor authentication: Before the period of access control evaluation, it is important to consider where to employ single or multi-factor authentication. For example, one-factor authentication may be sufficient for blood pressure or temperature readings, but accessing data from healthcare servers may require multi-factor authentication.
	        
	   \item Continuous authentication: Healthcare professionals need to use user credentials repeatedly throughout the day while accessing the patients' data. One existing solution is establishing a grace period after building the initial trust. However, the grace period may lead to malicious scenarios as any unauthorized personnel can access the device within the grace period. One possible solution is continuous authentication \cite{acar2018waca, mohaisen2013protecting, jeon2019privacy} that can be achieved by implementing different methods, such as wearable-assisted authentication, sensor-assisted authentication. 
	        
	   \item Emergency considerations: One needs to consider multiple scenarios for accessing the medical devices in the event that one method is not available. For example, a medical device may be considered only to transmit data after authentication, but still be allowed to access in any emergency scenario related to patient's health concerns~\cite{imprivata_2018}.
	   \end{enumerate}
	    
        
        
        

    \item \textit{Confidentiality:} Device information, system configuration, and healthcare data should be accessible only by authorized personnel or entities. These entities needed to be authenticated before accessing any healthcare-related confidential information. 
    However, it is possible to eavesdrop on existing healthcare devices, e.g., insulin pump communication channel and gets patients' data and device-related information \cite{li2011hijacking}.
    
    \item \textit{Integrity:} Data, device information, and system configuration should not be modifiable by unauthorized users' devices or applications. For instance, if there is no integrity checking mechanism in medical devices, data can be altered, and medical devices might accept malicious inputs, which can lead to severe attacks such as a code injection attack \cite{rieback2006your}. Current fitness tracker (e.g., Fitbit Charge, Garmin Vivosmart, etc.) devices lack integrity check mechanism for firmware updates \cite{ly2016security, clausing2015security}. 
    
    \item \textit{Non-repudiation:} A healthcare system performs different operations, and this information is usually kept secret in an access log. Any modification in this log should be traced and monitored and only performed by verified users. The attackers might want to delete these logs to cover their traces. For instance, the Fitbit smartwatch keeps their daily logs in clear-text files where an unauthorized user can change the log file to reverse engineer the communication protocol without keeping any trace \cite{rahman2013fit}. There are many resource-limited medical devices where there is no log in the systems, and attackers might try to gain access to the system without leaving any footprints.
    
    
    \item \textit{Availability:} The service provided by the healthcare system should always be available to the authorized users for accessing device systems, and patients' data in normal or emergency situations. For instance, an implementation flaw has been found in an ICD, which does not let the device go into the sleep mode when a communication session ends~\cite{dos}. Such a flaw can be exploited to trigger denial of service attacks, thus making the device unavailable.
    
\end{enumerate}

    
    In addition to satisfying the security goals mentioned above, one should ensure necessary privacy requirements in healthcare systems. 
    In this work, we consider the privacy goals based on device, data, and communication anonymity properties~\cite{avancha2012privacy, shahid2017ppvc, yuksel2017research}.

\begin{enumerate} [nosep, wide=0pt, leftmargin=*]

    \item \textit{Device anonymity}: It means that the identity of a medical device is unknown to the system so that unauthorized entities should not be able to determine the existence of the device type, specific device ID, and traditional identifiers such as IP and MAC addresses.
    
    \item \textit{Data anonymity}: The goal of data anonymity is to prevent unauthorized users from identifying a user and the user's sensitive data. Patients and doctors should not use their real identities; instead, they use pseudonyms or other temporary identifiers. 
    
    
    \item \textit{Communication anonymity:} Unauthorized entities should not be able to identify the connection between the user and the system. Effective necessary mechanisms, such as collision-resistant pseudonyms \cite{guo2014privacy} should be used to ensure anonymous communication.
    
    \item \textit{Unlinkability:} An attacker who tracks the data transactions between the sender and the receiver should not be able to establish a relationship between data and sender.
    
\end{enumerate}

\vspace{-0.4cm}
\section{Attack model and Existing security and privacy attacks on healthcare devices and applications}\label{threats}

    
As existing healthcare devices and applications fail to meet the security and privacy requirements (as discussed in Section~\ref{goals}), attackers can exploit different components of healthcare systems to perform malicious activities. In this section, we explain attack goals considering the capabilities of an attacker and attack methodologies to perform different attacks on healthcare systems. Moreover, we discuss various 
attacks on different components of healthcare systems (e.g., sensor, device, network, etc.) and summarize how attackers can compromise the security 
and privacy of targeted healthcare systems. We formally categorize existing security and privacy attacks on healthcare systems reported by the research community and developers and explain the attack methods and effects in detail.


\noindent\textbf{Attacker Goals:} An attacker can target a medical device to perform various malicious activities, including communication interception, data modification, device or data unavailability, etc. We categorize attack goals in the following categories based on the attack impacts on the healthcare devices and patients:

\begin{enumerate}[nosep, wide=0pt, leftmargin=*]

    \item \textbf{Hardware modification:} An attacker tries to tamper the device hardware architecture so that he/she can insert malicious hardware trojan during chip manufacturing time. 
    
    \item \textbf{Unavailability:} An attacker seeks to use malicious written programs to perform different attacks (e.g., malware, ransomware, etc.) and make the device and data unavailable (may be until a ransom is paid).
    
    \item \textbf{Communication delay:} An attacker attempts to connect with the healthcare device using an unauthorized programmer device (e.g., smartphone, personal computer, etc.) and forces the device to continue communication with an unauthorized programmer.
    
    \item \textbf{Data sniffing:} An attacker tries to capture the communication of healthcare devices' for collecting sensitive information such as the patient's vital signs and device information.
    
    \item \textbf{Data modification:} An attacker attempts to modify the patient's vital signs by breaking into the device or intercepting and modifying the communication packet between the healthcare device and the programmer device.
    
    \item \textbf{Information Leakage:} An attacker tries to retrieve confidential and sensitive information from healthcare systems. For instance, he/she can extract secret cryptographic keys, device power consumption, personal information (e.g., bank card, PINs, etc.) using several methods such as statistical analysis, EMI radiation, malicious software applications, etc.

\end{enumerate}

\noindent\textbf{Attacker Capabilities:} We consider following capabilities for an attacker to successfully implement different attacks on healthcare systems:

\begin{enumerate}[nosep, wide=0pt, leftmargin=*]
    
    \item An attacker has physical and/or remote access to healthcare systems environment.
    
    \item An attacker has the knowledge of which communication standard and protocol are used by the healthcare devices to establish communication with the programmer device.
    
    \item An attacker can access communication channels using third-party devices (e.g., sniffer, off-the-shelf hardware and software, etc.).
    
    \item An attacker can use a programmer device to impersonate her/himself as a patient to collect sensitive information from the healthcare device. 

\end{enumerate}

\noindent\textbf{Attack Types:} Depending upon their goals, capabilities, and relationship to the system, adversaries in healthcare systems can be categorized as follows:

\begin{enumerate} [nosep, wide=0pt, leftmargin=*]
    
    \item \textbf{Passive adversary:} An adversary of this kind can eavesdrop on communication channels, including side channels or unintentional communication channels without interrupting the communication. It is a direct threat to confidentiality and authentication for an insecure communication channel. By reading messages only, she/he may determine whether a person carries any medical device or not, find out device model, serial number, capture telemetry data, and disclose private information about a patient. Recently, the value of personal health information in underground markets has been rising. If no authentication mechanism is enforced, then the adversary can obtain private information (e.g., surgery type, social security number, etc.) related to the patient. 
    
    \item \textbf{Active adversary:} Such an adversary can interrupt the communication channel and read, modify, inject data. The adversary can be capable of capturing messages exchanged over the radio channel. The corresponding attacks may involve a sequence of interceptions, modifications, interruptions, and generations of extra messages to achieve different goals. Moreover, an active attacker can impersonate a programmer medical device (e.g., smartphone, personal computer, etc.), which is a third-party device used in IMD. 
    It can request confidential information, reprogram the medical device, cause a shock to the patient, or drain the battery of the medical device. An adversary may track a patient (e.g., patient's location, diagnosis, blackmail-worthy material, etc.) so that he can cause physical or psychological harm. 

\end{enumerate}

It is worth noting that it is not essential for an attacker to be close to the healthcare devices to conduct an attack. An adversary can be an external or internal entity with respect to the system. The adversary can also be a manufacturer, a patient, a physician, or even a hospital administrator. 

To understand the effect of the attacks on real-life healthcare systems, we introduce a vulnerability metrics based on a widely accepted measure of common vulnerability scoring system (CVSS) to quantify the impact of these attacks~\cite{mell2007complete}. We consider the following vulnerability metrics to illustrate the severity of different attacks on healthcare system:

    
\begin{table*}[t!]
\caption{List of existing attacks to healthcare devices and applications.}
\vspace{-0.5cm}
    \begin{tablenotes}
        \footnotesize
        \item [$\dagger$]Impacted security: integrity (I), availability (I), confidentiality (C).\\
        \item [$\dagger\dagger$]Vulnerability metrics: attack approach (AA), attack complexity (AC), privilege requirement (PR), user cooperation (UC).
    \end{tablenotes}
\begin{threeparttable}
\centering
\resizebox{1\textwidth}{!}{
\begin{tabular}{|c|c|c|c|c|c|c|c|c|c|}
\hline
\multirow{2}{*}{Attacks} & \multirow{2}{*}{Attack Type} & \multirow{2}{*}{\begin{tabular}[c]{@{}c@{}}Target Medical Devices\end{tabular}} & \multirow{2}{*}{\begin{tabular}[c]{@{}c@{}}Target\\ Component\end{tabular}} & \multicolumn{5}{c|}{\begin{tabular}[c]{@{}c@{}}Vulnerability\\ Metrics\tnote{$\dagger\dagger$}\end{tabular}} & \multirow{2}{*}{Ref.} \\ \cline{5-9}
 &  &  &  & AA & AC & PR & UC & Impact\tnote{$\dagger$} &  \\ \hline
\multicolumn{1}{|c|}{\multirow{1}{*}{\begin{tabular}[c]{@{}c@{}}Hardware\end{tabular}}}                             & Hardware Trojans                           & \begin{tabular}[c]{@{}c@{}} Active Therapeutic \\ Devices  \end{tabular}        & Sensor                                          & Active   & High   & -           & -           & I                      & \cite{wehbe2017novel, onwearfit, food2016postmarket, wehbe2018securing}                     \\ \hline
\multicolumn{1}{|c|}{\multirow{4}{*}{\begin{tabular}[c]{@{}c@{}}\\\\\\\\ Software\end{tabular}}}          & Malware                                          & \begin{tabular}[c]{@{}c@{}}Active Therapeutic \\ Devices\end{tabular}      & \begin{tabular}[c]{@{}c@{}}Device, \\Data,\\ Healthcare provider\end{tabular}                               & Active   & Low           & -           & \checkmark   & I, A            &  \cite{fu2014controlling, softwarethreats1, malware}                \\ \cline{2-10}   
                                      & Ransomware                                       & \begin{tabular}[c]{@{}c@{}}Active Therapeutic \\ Devices\end{tabular}    & \begin{tabular}[c]{@{}c@{}}Data,\\ Healthcare provider\end{tabular}                     & Active   & Low           & -           & \checkmark   & I, A              & \cite{martin2017cybersecurity, mansfield2016ransomware, ransomware, zep, life}                     \\ \cline{2-10} 
                                      & \begin{tabular}[c]{@{}c@{}}Outdated Operating \\Systems \end{tabular}   & \begin{tabular}[c]{@{}c@{}} Active Therapeutic \\ Devices \end{tabular}                       & \begin{tabular}[c]{@{}c@{}}Device, \\ Data,\\ Healthcare provider\end{tabular}                                  & Passive           & High   & -           & -           & I, A                & \cite{wan}, \cite{moses2015lack}                     \\ \cline{2-10}
                                      & \begin{tabular}[c]{@{}c@{}}Electroencepha\\-lography (EEG) \\ \end{tabular}                     & \begin{tabular}[c]{@{}c@{}} Non-invasive Devices \end{tabular}  & Device                 & Passive           & Low           & \checkmark   & -           & C      & \cite{martinovic2012feasibility}     \\ \cline{2-10} 
                                      & \begin{tabular}[c]{@{}c@{}}Counterfeit Firmware \\Update\end{tabular}                      & \begin{tabular}[c]{@{}c@{}}Invasive Devices, \\Non-invasive Devices\end{tabular}      & \begin{tabular}[c]{@{}c@{}}Data,\\ Healthcare provider\end{tabular}                      & Passive   & High   & \checkmark   & -           & I, A                 & \begin{tabular}[c]{@{}c@{}} \cite{hanna2011take}, \\\cite{rios2017security, rieck2016attacks, ly2016security, clausing2015security, kim2015burnfit, shim2017case, classen2018anatomy, arias2015privacy}    \end{tabular}   \\ \hline
                                      
 \multicolumn{1}{|c|}{\multirow{2}{*}{\begin{tabular}[c]{@{}c@{}}System-level\end{tabular}}}      & \begin{tabular}[c]{@{}c@{}}Weak Authentication \\Schemes \\ exploitations\end{tabular}                     &                                                    \begin{tabular}[c]{@{}c@{}} Invasive Devices, Non-invasive \\Devices, Active Therapeutic \\Devices\end{tabular}        & \begin{tabular}[c]{@{}c@{}}Device, \\Data,\\                                         Healthcare provider\end{tabular}                           & Passive   & High   & \checkmark   & -           & C, I              & \begin{tabular}[c]{@{}c@{}} \cite{hanna2011take,moses2015lack},  \\\cite{ xiao2019can, radio8, bd, aestiva, mcmahon2017assessing, change, medt, philips, mahler2018know, natus, rahman2013fit} \end{tabular}                    \\ \cline{2-10} 
                                      & Privilege Escalation                             & Invasive Devices                 & \begin{tabular}[c]{@{}c@{}}Device,\\ Data\end{tabular}                                   & Passive   & Low           & \checkmark   & \checkmark   & I, A               & \cite{yan2014semantic}   \\\hline                
\multicolumn{1}{|c|}{\multirow{4}{*}{\begin{tabular}[c]{@{}c@{}}Side-channel\end{tabular}}}         & \begin{tabular}[c]{@{}c@{}}Electromagnetic \\Interference\end{tabular}                & \begin{tabular}[c]{@{}c@{}}\textit{Invasive Devices}\end{tabular}            & Sensor                                            & Passive   & High   & -           & -           & A               & \cite{kune2013ghost, hayes1997interference, jilek2010safety}                     \\ \cline{2-10} 
                                      & \begin{tabular}[c]{@{}c@{}} Sensor \\Spoofing     \end{tabular}                             & Invasive Devices   &               Sensor                    & Active   & High   & -           & -           & A                 & \cite{park2016ain}  \\ \cline{2-10} 
                                      & \begin{tabular}[c]{@{}c@{}} Differential Power \\Analysis\end{tabular}   & Non-invasive Devices       &               Device                    & Passive   & High   & -           & -           & I, A                 & \cite{zhang2013towards} \\ \hline
\multicolumn{1}{|c|}{\multirow{12}{*}{\begin{tabular}[c]{@{}c@{}}Communication \\Channel\end{tabular}}}  & Eavesdropping                                    & \begin{tabular}[c]{@{}c@{}}Invasive Devices, \\Non-invasive Devices\end{tabular}                     & Network                                   & Passive   & Low   & -           & -           & C, I              & \cite{li2011hijacking}, \cite{blindattack, radio5, article, cusack2017assessment, marin2016security, bonaci2014securing, bonaci2014app, zhang2017security, kim2015vibration, fawaz2016protecting, halevi2010pairing, lotfy2016assessing, wood2017cleartext, li2015brain}                     \\ \cline{2-10} 
                                      & Replay                                    & \begin{tabular}[c]{@{}c@{}}Non-invasive Devices\end{tabular}                        & Network               & Active   & Low   & -           & -           & C, I            & \begin{tabular}[c]{@{}c@{}} \cite{radio5}, \cite{radcliffe2011hacking}, \\\cite{xiao2019can}  \end{tabular}                   \\ \cline{2-10} 
                                      & Impersonation                             & \begin{tabular}[c]{@{}c@{}}Non-invasive Devices\end{tabular}                                          & Network              & Passive   & High   & \checkmark   & -           & I                         & \cite{li2011hijacking}, \cite{flynn2020knock}                     \\ \cline{2-10} 
                                      & Denial-of-service                  & \begin{tabular}[c]{@{}c@{}}Invasive Devices, Non-invasive \\ Devices,  Active Therapeutic \\Devices\end{tabular} & Network & Active   & Low          & \checkmark   & -           & A                & \begin{tabular}[c]{@{}c@{}}\cite{halperin2008pacemakers}, \\\cite{dos, ransford2017cybersecurity, twelve, alsubaei2017security, wang2019security}, \\\cite{natus}, \cite{mcmahon2017assessing}     \end{tabular}                \\ \cline{2-10} 
                                      & \begin{tabular}[c]{@{}c@{}}Multiple Input and \\Multiple Output\end{tabular} & \textit{Invasive Devices}                                         & Device                       & Passive   & High   & -           & -           & C          & \cite{tippenhauer2013limitations}                     \\ \cline{2-10} 
                                      & \begin{tabular}[c]{@{}c@{}}Man-in-the-middle\end{tabular}                         & \begin{tabular}[c]{@{}c@{}}Invasive Devices, \\Non-invasive Devices \end{tabular}   & Network     & Active   & High   & -           & -           & C, I    & \begin{tabular}[c]{@{}c@{}} \cite{fawaz2016protecting}, \\\cite{cremers2012distance,pournaghshband2012securing, hei2014patient, chauhan2016characterization, paoletti2019synthesizing}  \end{tabular}                   \\ \cline{2-10} 
                                      & Battery depletion                         & \begin{tabular}[c]{@{}c@{}}Invasive Devices, \\Non-invasive Devices\end{tabular} & Device & Active   & Low           & -   & -           & A                  & \begin{tabular}[c]{@{}c@{}}\cite{dos, ransford2017cybersecurity}, \\\cite{raymond2009effects, hei2013security, safavi2014improving}      \end{tabular}               \\ \hline
\end{tabular}}
    \end{threeparttable}
    \label{table:Attacks_list}
        \vspace{-0.5cm}
\end{table*}
      

\begin{enumerate}  [nosep, wide=0pt, leftmargin=*]
        
    \item \textbf{Attack approach (AA):} It reflects how an attacker exploits a healthcare system to perform malicious activities. Based on the attack approaches, it can be categorized as follows: active attack and passive attack. Passive attacks refer to an attack that performs malicious activities in a healthcare system without obstructing the normal operation of the system, whereas active attacks obstruct the normal operation of a healthcare system to perform malicious activities.
        
    \item \textbf{Attack complexity (AC)}: This metric specifies the amount of information an adversary needs to perform an attack on a healthcare system.  An attacker needs partial (e.g., device model, used communication protocol, etc.) or full information (e.g., network structure, encryption type, etc.) of healthcare devices to perform an attack. For instance, a man-in-the-middle attack needs physical access (high complexity) to the network, whereas a replay attack can be performed by capturing the communication packet passively (low complexity) and sending the same packet repeatedly. 
        
    \item \textbf{Privilege requirement (PR)}: To perform an attack, the attacker needs certain privileges or access to the healthcare system. We use the required privilege of an attacker to the system to explain the impact of the attack. For example, a communication medium attack such as packet sniffing does not need any access to perform malicious activities while impersonation attack requires access to the healthcare system.
        
    \item \textbf{User cooperation (UC)}: An attack may require human interaction other than the attacker to exploit the vulnerability successfully. For instance, to install malware, user interaction is needed in the healthcare system.
        
\end{enumerate}

In the following subsections,  we group the existing attacks according to their relevant attack surfaces and provide an explicit categorization of the attacks. We also present a detailed summary 
in Table~\ref{table:Attacks_list}.

\vspace{-0.3cm}    
\subsection{Hardware Attacks}
    
Hardware attacks refer to an exploitable weakness in a device hardware that can be used to gain physical or remote access to the device to perform malicious activities. An attacker may know or gain access to the internal hardware architecture of the device and insert \textit{hardware trojans} (HTs) during chip manufacturing that can lead to data corruption, causing serious harm to the medical devices~\cite{wehbe2017novel}. Indeed, HTs have emerged as a major security concern for integrated circuits (ICs) as most ICs are manufactured in outsourced fabrication facilities. Third-party vendors can include unverified intellectual property cores that act as HTs and can be used to perform malicious activities, including leaking information from the medical devices. HTs can be classified based on \textit{physical attributes} (e.g., chip layout, activation, etc.) and \textit{action characteristics} (e.g., logic functions, chip activities, etc.) \cite{tehranipoor2010survey}. \textit{Physical attributes} describe the trojans that can be injected through the addition or deletion of transistors or gates in the chip manufacturing stage. \textit{Action characteristics} refer to an HT where it changes the chip's function by adding or bypassing existing logic.

The 
FDA has released numerous reports on changing patient's health data by modifying medical device hardware \cite{onwearfit, food2016postmarket}. In recent work, researchers presented an HT attack on the bacillus calmette guerin (BCG) scale~\cite{wehbe2018securing}. They injected a malicious payload that modifies the logic of an XOR gate on the input bus. It is a relatively less severe HT attack and cannot be detected if spread out among thousands of gates in the embedded system of the healthcare device.

    
\vspace{-0.3cm}    
\subsection{Software Attacks}
     
     
Software attacks refer to maliciously written programs to deliberately impact healthcare devices, associated computers, or servers. The use of embedded and customizable software in a healthcare environment is increasing rapidly, which certainly improves the patients' treatment and monitoring. However, there is no satisfactory security measure to verify the functionalities and authenticity of medical software. As a result, healthcare systems are facing various threats related to software and apps such as \textit{malware, ransomware, outdated operating systems, counterfeit firmware update, and electroencephalography (EEG)} attacks.
     
        
\vspace{4pt}
\noindent\textbf{Malware:} Any software or application that is written with malicious intent is called malware. A healthcare device infected with malware can stray away from its normal functionalities such as slow or shut down a device. For instance, \textit{Conflicker}, a relatively old malware, was recently detected on 104 devices, including X-ray machines, mammography, and a gamma camera for nuclear medicine at the James A. Haley Veteran's Hospital in Tampa, Florida, USA \cite{fu2014controlling}. This malware affected Microsoft Windows operating system from a thumb drive because the network drivers were not patched with the \textit{MS08-067} patch from Microsoft. Hence, a remote and unauthenticated attacker could execute arbitrary code on the vulnerable system. In January 2010, a veterans affairs (VA) catheterization laboratory in New Jersey, was temporarily closed due to a malware infection into the computer systems \cite{softwarethreats1}. As a result, patients were unable to get any medical services from that hospital. Affected devices were X-ray machines and lab equipment manufactured by reputed companies. Moreover, malware like \textit{Kwampirs} can introduce instability into healthcare systems by triggering equipment malfunction or delay in accessing information \cite{malware}.
         
\vspace{4pt}
\noindent\textbf{Ransomware:} Ransomware is a unique subset of malware that limits or blocks users' access by locking the system and data unless a ransom is paid. In May 2017, around 50 hospitals in the U.K. were directly affected, and many hospitals preemptively shut down their computer systems due to ransomware. It caused considerable disruption, such as affecting care delivery, compromising patient safety, and potentially eroding trust~\cite{martin2017cybersecurity}. The ransomware encrypted and blocked the patient's data and threatened to publish or delete them unless a ransom is paid. In 2016, a ransomware shut down the network of the Hollywood Presbyterian Medical Center in Los Angeles, California for ten days. It prevented its staffs from accessing medical records or using medical equipment until the hospital paid a ransom of \$17,000 \cite{mansfield2016ransomware}. Freedom of information requests in the U.K. found that in 2015-16 up to half of the national health service (NHS) trusts were hit by ransomware \cite{mansfield2016ransomware}. Also, two US-based health centers (Hancock Health and Erie County Medical Center) were hit by \textit{SamSam} ransomware and ended up meeting the ransom demand. In all these incidents, on average, it would take 12 days to restore limited system access and six weeks to restore full access to the system~\cite{ransomware}. 

Recently, a new ransomware called \textit{Zeppelin} has been reported in healthcare companies across Europe, USA, and Canada \cite{zep}. \textit{Zeppelin} is a Delphi-based highly-configurable ransomware that could be deployed as an \textit{.exe}, \textit{.dll}, or wrapped in a \textit{PowerShell loader}. This ransomware employed a standard combination of symmetric and asymmetric encryption with randomly generated keys for each file (advanced encryption standard (AES)-256 in cipher-block chaining (CBC) mode). 

\vspace{4pt}
\noindent\textbf{Outdated operating systems:} Outdated operating system (OS) poses severe threats to healthcare devices as new-found bugs are not addressed in the older versions of the OS by the vendor. As a result, attackers can exploit the existing bugs of the OSes by simply injecting malicious code snippets or software. According to the Duo security research team, 70\% of healthcare devices in North America and Europe will still be running outdated Windows 7 OS at the end of 2020, although Microsoft stopped releasing any patches for Windows 7~\cite{wan}. As an example, the \textit{WannaCry} ransomware attacks were launched against unpatched healthcare devices, where IT professionals neglected to download the OS update on time. A group of researchers conducted a vulnerability assessment in a radiology department where the majority of the networked medical equipments (e.g., medical ventilator, X-ray machine, anesthetic machine, etc.) were running on an old and insecure version of OSes (e.g., Windows Vista, Windows XP, etc.) \cite{moses2015lack}. These OSes were running unprotected, insecure, and vulnerable applications that had no firewall or protection against malware.

\vspace{3pt}
\noindent\textbf{Counterfeit firmware update:} Counterfeit firmware in medical devices introduces numerous threats to healthcare devices as an attacker can gain access to the devices and manipulate the applications using fake copies of firmware. Counterfeit firmware is produced and distributed in such a way that it appears to be authentic. Hanna et al. analyzed an automated external defibrillator (AED) (Cardiac Science G3 Plus 9390A) and found four vulnerabilities, including a software update mechanism that accepts counterfeit firmware~\cite{hanna2011take}. An attacker could replace the firmware of that specific AED with custom firmware designed to exploit the \textit{AEDUpdate package} to perform buffer overflow. Rios et al. showed that it is possible to update unverified firmware of a home monitoring device that is connected to an ICD~\cite{rios2017security}. As there was no digitally signed firmware within the ICD ecosystem, the unverified firmware allowed to perform a man-in-the-middle attack on the ICD. Rieck et al. reverse engineered a wearable healthcare device (Withings Activite fitness tracker) to identify and reconstruct the header structure within a firmware update~\cite{rieck2016attacks}. As the authentication scheme of this device only computes a checksum over the actual content of the image, it is possible to create a fake copy of the firmware by alternating the checksum in the header field. In other studies~\cite{ly2016security, clausing2015security, kim2015burnfit, shim2017case, classen2018anatomy, arias2015privacy}, fitness tracker devices' application codes and firmware were reverse engineered to extract and manipulate fitness-related data. These devices lacked authentication, encryption, and integrity check mechanisms for firmware updates.

\vspace{3pt}
\noindent\textbf{Electroencephalography (EEG) attacks:} In an EEG-based attack, an attacker is a malicious third-party application developer who is using an EEG-based brain-computer interface (BCI) device. The main goal of this device is to learn secret and private information about the user. An attacker developed a malicious software called \textit{brain spyware} that was integrated into a BCI device to detect private information 
of the user \cite{martinovic2012feasibility}. Moreover, an attacker can specially design the videos and images that can be shown to the user to maximize the information leakage from the BCI device during the time of the attacks. It has been shown that the captured electroencephalography (EEG) signal can reveal personal information (e.g., bank cards, PINs, area of living, etc.). In other studies \cite{bonaci2014securing, bonaci2014app}, researchers used brain spyware to extract not only private information about users' memories and prejudices, but also their possible neurological disorders. 

\vspace{-0.3cm}
\subsection{System-level Attacks}

This type of attack directly focuses on system-level vulnerabilities such as memory modules, system applications, and design flaws in a healthcare system. Attackers can exploit these vulnerabilities to gain unauthorized control and access to sensitive data. There are two major types of system-level attacks that can be performed on healthcare system. These are \textit{weak authentication scheme exploitations} and \textit{privilege escalation attacks} on healthcare devices.

\vspace{4pt}
\noindent\textbf{Weak authentication scheme exploitations:} Authentication is a process where one needs to prove his/her identity to an application or system to access a service. Weak authentication describes a scenario where the strength of the authentication mechanism is relatively weak compared to the value of the assets. In a recent study, researchers investigated weak password-based authentication in healthcare devices focusing on external and internal defibrillators. According to the study, an AED using \textit{MDLink} software has a weak password authentication scheme where the password file is stored on the local hard drive \cite{hanna2011take}. As a result, anyone with privileges could delete or change the password file and install any new software on the machine. Furthermore, researchers reverse engineered the MDLink authentication mechanism and wrote a small utility to change or recover a user's password. In another work, Xiao et al. used a malicious brain-computer interface (BCI) app to steal the patient's EEG data by exploiting the standard software development kit (SDK) as the calling application programming interfaces (APIs) have no authentication schemes~\cite{xiao2019can}. In 2017, a group of researchers identified a hard-coded authentication system in the Medfusion 4000 Wireless Syringe Infusion Pump from Smiths Medical that allowed FTP server connections without any verification~\cite{radio8}. In a recent article, researchers reported an EEG-based application that allows to execute malicious code on the EEG device. When a client requests an EEG file, this application exploits the path requested by the client using the buffer overflow to remotely access the EEG device and make the device unavailable~\cite{natus}.

Most networked medical devices pose weak authentication schemes during the time of reading or writing data from these devices. Moses et al. conducted a study on onsite networked medical devices in a radiology department to identify vulnerabilities by using a port scanner and a network vulnerability scanner~\cite{moses2015lack}. This study reported that around 85\% of the networked medical devices allowed unauthorized users to read or write data from a portable USB storage medium. Moreover, a CD or DVD drive in 17 out of 31 networked items allowed unauthorized users to copy or upload data from the equipment. Researchers from \textit{CyberMDX} studied the improper authentication vulnerabilities in \textit{GE Aestiva} and \textit{Aespire Anesthesia} devices \cite{aestiva}. In this work, researchers used serial devices to connect to a TCP/IP server via an unsecured terminal that allowed remote access to modify device configuration and disable alarms. Security researchers of \textit{Alfonso Powers} and \textit{Bradley Shubin} studied connected cardiology devices made by Change Healthcare \cite{change}. They have reported insecure file permission in the default installation that might allow an attacker with local system access to execute unauthorized arbitrary code. Security researchers of Philips reported improper authentication and missing sensitive data encryption vulnerabilities in medical image management systems that could enable an attacker to see usernames, passwords, and personal data \cite{philips}. In some cases, weak authentication in the image management allowed direct access to the memory locations to execute arbitrary code, alter the intended control flow, or cause the system to crash. Mahler et al. reported authentication flaws in medical imaging devices such as magnetic resonance imaging (MRI) or computed tomography (CT) machines. The scan configuration file inside a host controller PC of imaging device used default username and password that allowed an attacker to manipulate the file to change the CT's behavior or control the entire CT operation \cite{mahler2018know}. 

Several Medtronic ICDs and associated equipment use conexus radio frequency protocol that does not have any authentication or authorization~\cite{medt}. This allows a nearby attacker to access any implantable cardiac devices with radio turned on. An attacker can inject, replay, modify, and/or intercept data within the telemetry communication using unauthorized access. Rahman et al. reported that the Windows supported Fitbit application stored its daily logs containing data requests, responses, and social network data in clear-text files without any authentication scheme~\cite{rahman2013fit}. Researchers reverse engineered the communication protocol (ANT) of Fitbit and demonstrated active and passive attacks using the off-the-shelf software module.  

Researchers of CyberMDX also discovered two security vulnerabilities in the firmware and web management of Alaris Gateway Workstations (AGWs) that were used to provide mounting, power, and communication support to infusion pumps \cite{bd}. AGWs were vulnerable to an exploit where an attacker could remotely exploit firmware files, which required no special privileges to execute. Moreover, an attacker could manipulate gateway communication with connected infusion pumps. As the web management system did not require any credentials or passwords, attackers could easily connect to a workstation using IP address and monitor infusion pump's status, event logs, etc. Mcmahon et al. showed that an earlier version of Dropbear SSH Server used by several IMDs provided memory access to any users without any proper authentication~\cite{mcmahon2017assessing}. An attacker could get local access to the process memory by simply running trace with -v option that might disclose sensitive information of the patient held on the database.

\vspace{4pt}
\noindent\textbf{Privilege escalation attacks:} A privilege escalation attack takes advantage of bugs, design flaws, or configuration failures in an OS or application to access healthcare devices and data that usually require exclusive permission or authorization. Privilege escalation attacks can be launched by rogue users (e.g., patients or physicians) who have access to healthcare systems and perform malicious activities, such as calibration failures, data modification, etc. Yan et al. introduced two types of privilege escalation attacks (\textit{pressure-based} attacks and \textit{time-based} attacks) to IMDs, which can mislead the diagnostic process by altering collected and stored data after bypassing the initial access control mechanism \cite{yan2014semantic}. In pressure-based attacks, the attacker can change the pressure value of the sensors connected to the IMDs to report misdiagnosis of the patient. The attacker postpones the pressure data of some pressure sensors for certain time slots in time-based attacks.


\vspace{-0.3cm}     
\subsection{Side-channel Attacks} 
    
Side-channel attacks aim at extracting sensitive data (on-going task, encryption method, etc.) from a healthcare system/device by analyzing physical parameters without interrupting the on-going task. 
Examples of physical parameters include how the circuit works, what data it is processing, and when a victim's device is being used, etc. There are three major types of side-channel attacks on healthcare systems. These are \textit{electromagnetic interference, sensor spoofing, and differential power analysis} attacks on medical devices.
    
\vspace{4pt}
\noindent\textbf{Electromagnetic Interference (EMI) attacks:} EMI attacks are performed by measuring the electromagnetic (EM) radiation emitted from a device and performing signal analysis to infer sensitive information. Kune et al. showed that analog sensors used in medical devices (e.g., infusion pump, ICD, etc.) are sensitive to EMI and can provide an unchecked entry point into the medical devices~\cite{kune2013ghost}. Here, an attacker can inject EMI signals in an ICD's sensing unit to alter the sensor readings and trick the medical devices to prevent data communication. Additionally, several prior works reported that EMI could cause device malfunction in pacemakers and ICDs~\cite{hayes1997interference, jilek2010safety}. 
    

\vspace{4pt}
\noindent\textbf{Sensor spoofing attacks:} In a sensor spoofing attack, an adversary alters the physical environment in a way so that a medical system behaves abruptly~\cite{sikder2019aegis, sikder2019context}. Park et al. introduced a sensor spoofing attack against the infrared drop (ID) sensor embedded in the infusion pump \cite{park2016ain}. An ID sensor has a linear property of input-output stimuli, which can be manipulated to non-linear behavior by exceeding the upper bound operating region of the infusion pump. Researchers showed that an attacker could inject an external power signal to a targeted sensor to block the sensor response to environmental changes, which results in over-infusion or under-infusion to the patient. Over-infusion allowed the infusion pump to infuse about 333\% of the fluid as compared to the normal operation while under-infusion infused approximately 45\% less than the normal operation.

\vspace{4pt}
\noindent\textbf{Differential power analysis (DPA) attacks:} DPA attacks use different analysis techniques (statistical, error correction, etc.) to infer sensitive information from power consumption data. Zhang et al. introduced a DPA attack that can extract secret keys from extremely noisy channels in a heart-rate monitor using symmetric cipher~\cite{zhang2013towards}. Here, the heart-rate monitor uses AES encryption to encrypt the measured heart rates before transmitting to an end device (hub or storage). An attacker can recover the secret key used in the encryption scheme by analyzing the current consumption rate while measuring the heart rate of the patients. If the same key is used in the same model of all heart-rate monitor devices, an attacker can publicize the inferred secret key and thus make the cryptographic protection ineffective for a large number of devices.

\vspace{-0.3cm}    
\subsection{Attacks via Communication Channel}
 
Wireless communication is used for the connectivity among healthcare devices for remote monitoring, diagnosis, treatment, and emergency support. For healthcare devices, attacks through communication channels have become a major concern as attackers can perform various attacks including \textit{eavesdropping, replay, impersonation, denial-of-service, multiple input and multiple output, man-in-the-middle, and battery depletion attacks} to compromise the integrity of the device operation.
    
    
\vspace{4pt}
\noindent\textbf{Eavesdropping:} Eavesdropping refers to an attack where an adversary tries to steal information over a communication medium by taking advantage of the unsecured communication channels. Several eavesdropping attacks on medical devices (e.g., Medtronic and OneTouch Ping insulin pumps) have been reported, which captures the clear text communication to capture sensitive patient data such as blood glucose results and insulin dosage~\cite{blindattack, radio5}. Li et al. demonstrated an eavesdropping attack in an insulin pump by using off-the-shelf hardware and a software radio platform \cite{li2011hijacking}. As the communication channel does not use any authentication, researchers showed that attackers could capture glucose level, device type, device PIN, and medical condition of the patient by eavesdropping the communication channel. A group of researchers was able to capture enough sensitive data from Withings Blood Pressure Monitors' network traffic to determine the time and frequency of blood pressure testing on a patient~\cite{wood2017cleartext}. As the information sections of all queries and responses in Withings devices are transmitted in clear-text format, an attacker can easily monitor and capture network traffic to eavesdrop sensitive data including device ID, device type, and patient's readings.

Wearable healthcare devices, such as smartwatches and fitness bands, are also vulnerable to passive eavesdropping attacks as attackers can capture network traffic and sensitive data by simply using a sniffer module. Cusack et al. used a BLE sniffer to capture communication packets of four wearable devices using BLE 4.0 and 4.2 (e.g., Fitbit Charge HR, Samsung Gear3, etc.) and performed packet analysis to extract sensitive information~\cite{cusack2017assessment}. The captured packets were uploaded to \textit{Wireshark} for further analysis, and researchers founded that sensitive information, such as the long-term key to the BLE pairing process, sender, and receiver mac address, 
and communication messages were transmitted as plain text. 
Two wearable smartwatches' (TW64 and Mambo HR) traffic were remotely sniffed and analyzed using TI SmartRF and BLETestTool~\cite{zhang2017security}. Attackers could remotely control these two smartwatches, e.g., made them vibrating for a long time, possibly by keep sending fake command messages. These two devices had no technical security protection mechanism at all. 
Fawaz et al. performed a passive eavesdropping attack on BLE-enabled healthcare devices by sniffing the communication over advertisement channels~\cite{fawaz2016protecting}. Further analysis of these captured packets revealed that BLE-enabled healthcare devices use a fixed bluetooth address for long periods making the address randomization process ineffective. Hence, an attacker can sniff and capture sensitive health data of a patient without any interruption for a long period. Furthermore, the authors recovered the original Bluetooth signal from the jammed signal using a multiple input and multiple output receiver that contains detailed information of the patient's vitals. Lofty et al. captured the network traffic between a smartwatch and smartphone and showed that it is possible to convert the HEX-encoded data to human-readable data using reverse engineering techniques~\cite{lotfy2016assessing}. 
A passive attack was accomplished to sniff the internal LAN on an infusion pump, which was integrated into the IT networks \cite{article}. In this work, researchers found an open port in the infusion pump unit where the default password setting was not changed. Moreover, the information on the correct login was not monitored, and the communication was unencrypted.

Kim et al. showed that it is possible to infer the encryption key by capturing the vibration of a smartphone while transmitting to an IWMD~\cite{kim2015vibration}. This vibration of the smartphone also leaked an audible acoustic signal that was captured using a microphone. The recorded acoustic signal was highly correlated to the vibration waveform that could effectively block the transmission of the encryption key. Halevi et al. reported that auxiliary audio channels could be breached by close-range eavesdropping \cite{halevi2010pairing}. Here, researchers eavesdropped on the IMD key pairing process and detected the initial sequence of the secret key using a signal processing algorithm. Moreover, they extracted spectrum features from each consecutive bit and used these features as input to machine learning algorithms for classifying each bit value. Li et al. showed that the communication between prosthetic limb application and neural implant devices could be eavesdropped to capture brain neural signals, decompose raw signals, and obtain users' private information~\cite{li2015brain}. Furthermore, an attacker can get control of prosthetic limbs of patients and give dangerous movement to patients without being in the close proximity of the victim.


\vspace{3pt}
\noindent\textbf{Replay attacks:} A replay attack is a form of attack in which an adversary intercepts the data transmissions and fraudulently re-transmits it to misdirect the receiver. For instance, One touch Ping insulin pumps and blood glucose meters do not use any sequence numbers or timestamps, which allows attackers to capture transmissions and replays them later to perform an insulin bolus without specialized knowledge \cite{radio5}. A researcher, Jerome Radcliffe, showed that a continuous glucose monitoring device (CGM) without any timestamp or other protection methods in network packets could be exploited by a replay attack~\cite{radcliffe2011hacking}. This attack led to an unusual insulin dosage to the patient resulting in the hypoglycemic condition. In another work, Xiao et al. showed that software-defined radio (SDR) waves emitted from EEG devices can be recorded to replay in an RF dongle to recover the patient's EEG signals maliciously~\cite{xiao2019can}.

\vspace{3pt}
\noindent\textbf{Impersonation attacks:} In an impersonation attack, an adversary successfully disguises as a valid user in the communication system to gain access to the victim's sensitive information and take advantage of the clear text communication between healthcare devices. Li et al. showed that an unencrypted communication between a glucose monitoring device and the insulin delivery system could be sniffed, and by applying reverse engineering methods, it is possible to discover the device PIN~\cite{li2011hijacking}. Furthermore, this PIN can be used to authenticate a patient maliciously to perform an impersonation attack. In another work, researchers introduced a \textit{hijacking attack} using a smartphone application and its corresponding Medical IoT (MIoT) devices (e.g., pulse oximeter, glucometer, etc.)~\cite{flynn2020knock}. MIoT devices can store offline readings when the user's smartphone application is not available to upload the results in the smartphone interface. A hijacker with stolen user credentials can open an account from another smartphone and retrieve all the offline readings that can be verified using manual and digital forensics techniques~\cite{babun2018iotdots}.
        
\vspace{3pt}
\noindent\textbf{Denial-of-service (DoS) attacks:} In DoS attacks, the attackers usually make a healthcare device or system unavailable temporarily or permanently to the legitimate users by sending excessive and unnecessary service requests. For example, an ICD remains in the standby mode for 5 minutes after activation even though there is no active communication session. This wait time can be exploited by initiating false communication sessions and keep the ICD in standby mode for longer times \cite{dos}. Ransford et al. 
reported a \textit{crash attack} to cardiovascular implantable electronic devices (CIEDs) in which attackers send undisclosed radio traffic to disrupt the radio connectivity of CIED, causing the device to stop working \cite{ransford2017cybersecurity}. A group of researchers reverse engineered the communication protocol of a battery-powered ICD to make a communication with an unauthenticated device that posed a potential DoS risk to the ICD \cite{halperin2008pacemakers}. An exploitable DoS vulnerability was identified in the use of a return value in an EEG-based software applications program \cite{natus}. As a consequence, a specially crafted network packet could cause an out of bounds read to trigger this vulnerability.

Communication protocols of healthcare devices are also vulnerable to DoS attacks. A team of cybersecurity researchers reported \textit{SweynTooth}, a repository of twelve security vulnerabilities, affecting thousands of BLE-enabled smart medical devices \cite{twelve}. This repository includes a DoS attack where an attacker in radio range performs a buffer overflow by manipulating the link-layer length field. This attack triggers a deadlock state when a device received a packet with a clear link layer ID, primarily leading to an OS attack. Wang et al. \cite{wang2019security} reported a data overflow vulnerability in the medical image-based communication standard called digital imaging and communications in medicine (DICOM). Researchers developed a DICOM vulnerability framework and found that when the content of the received image file was greater than 7080 lines, archiving and communication systems refused to respond to any request from the server.

Medical web servers are also being targeted by the attackers to perform DoS attacks by sending numerous fake requests. In 2014, one of the largest children's hospitals in the USA was the target of a distributed DoS attack by flooding the website with numerous fake requests over a seven day period~\cite{alsubaei2017security}. In consequence, the hospital's website was unreachable and day-to-day operations at the hospital were slowed down. In another work, mcmahon et al. exploited an outdated versions of hypertext preprocessor (PHP < 4.4.5) to perform DoS attacks, as well as remote code executions~\cite{mcmahon2017assessing}.
        
\vspace{3pt}
\noindent\textbf{Multiple Input and Multiple Output (MIMO) attacks:} MIMO refers to a setting where multiple antennas used by the transmitter to transmit a wireless message to a receiver with multiple antennas. In the MIMO attack, attackers try to recover signals sent by the transmitter in the presence of a friendly jammer, without any collaboration with the jammer or transmitter. An attacker can recover confidential messages from distances even when the friendly jammer and the data source are few centimeters apart, and the attacker is several meters away. Friendly jamming is often used to protect the confidentiality of the communicated data, which also enables message authentication and access control. Researchers showed that MIMO attacks are still possible with two receiving antennas from a range up to 3 meters \cite{tippenhauer2013limitations}.
        
\vspace{3pt}
\noindent\textbf{Man-in-the-middle (MITM) attacks:} A MITM attack occurs when communication between different components of healthcare systems is monitored and modified by unauthorized users. This attack can be used to inject malicious codes to a healthcare device or server, intercept sensitive information like protected health information, expose confidential information, and modify trusted information. Researchers introduced \textit{distance hijacking} attack to intercept an ongoing communication in healthcare systems~\cite{cremers2012distance}. Here, two medical devices, \textit{prover} and \textit{verifier}, are considered in the distance bounding protocol, where verifier establishes physical proximity with the prover. The authors considered various adversarial capabilities for falsifying physical abilities to the prover to create a false or rogue prover that can intercept the communication and establish a new communication channel with the authorized verifier. In another study, a Bluetooth-enabled medical device named pulse oximeter was used to perform MITM attack~\cite{pournaghshband2012securing}. In this attack, attackers jammed the Bluetooth device to break the existing connection to pair the device with an access point (AP). Hei et al. presented a MITM attack where an attacker compromised the wireless communication between an insulin pump and a USB device~\cite{hei2014patient}. As the communication in the insulin pump was unencrypted, the attacker could perform \textit{signal acute overdose} and \textit{chronic overdose attacks}. \textit{Signal acute overdose} issued a one-time overdose to the patient, whereas the \textit{chronic overdose} issued extra portions of medication to the patient over a long period of time. Marin et al. demonstrated a MITM attack by intercepting the communication between the ICD and the programmer device~\cite{marin2016security}. Additionally, the researchers reverse engineered the communication protocol of the ICD to show that these proprietary protocols do not provide any security during communication. 

Fawaz et al. demonstrated a MITM attack against BLE-enabled healthcare devices that accept connection from unauthorized programmer devices~\cite{fawaz2016protecting}. The unauthorized BLE pairing allows an active attacker to inject malicious traffic into any BLE-channel at any given point of time without crossing the bounds of BLE specifications. Hence, an attacker can obtain an inventory of the patient's device and learn the patient's health condition, preferences, habits, etc. Chauhan et al. used a MITM proxy to capture and decrypt the network traffic generated by the smartwatch apps~\cite{chauhan2016characterization}. Researchers inspected the captured traffic and found personal information about the user, such as location, app credentials, health data (e.g., heart rate, water intake, etc.), and user activities as a result of the unencrypted communication between the device and app. Palotti et al. presented a formal approach to perform an effective and stealthy \textit{reprogramming attacks} on ICDs \cite{paoletti2019synthesizing}. Researchers focused on the ICD software that implements \textit{discrimination algorithm} along with multiple discrimination criteria (discriminators) for the detection and classification of arrhythmia episodes based on the analysis of intracardiac signals features. In this attack, an attacker tried to change the discrimination features that might alter the device's parameter to induce misclassification and inappropriate or missed therapy to the patient. For performing this attack in real-life, an attacker needs to know the ICD model of the victim so that it can select the appropriate discrimination algorithm. In addition to this work, an attacker can also send discovery signals to the device to know the ICD model~\cite{halperin2008pacemakers}.
        
        
\vspace{4pt}
\noindent\textbf{Battery depletion attacks:} A battery depletion attack is a forced authentication attack where an attacker tries to connect with an IMD to perform multiple authentications and drain the battery of the device. Raymond et al. presented a \textit{denial-of-sleep attack} that prevents the medical device to activate power-down mode in case of failed authentication attempt to exhaust the battery life~\cite {raymond2009effects}. Security researchers of MedSec studied St. Jude Medical Merline's CIED and reported a battery drain attack that reduces the CIED operating time cycle \cite{ransford2017cybersecurity}. In a recent report, an implementation flaw in an ICD is reported where the ICD does not go to the sleep mode even after ending an active communication session~\cite{dos}. This flaw can trigger a DoS attack and drains the battery of the ICD. Hei et al. presented a battery depletion attack on the IMD by exploiting the wireless communication between the IMD and programmer device~\cite{hei2013security}. As the programmer device needs to authenticate itself to the IMDs, an unauthorized programmer device can send several authentication requests to consume a considerable amount of battery life of the IMD. Researchers reported unsecured authentication in earlier versions of wearable devices, such as Google Glass, which allows root access to the attackers~\cite{safavi2014improving}. Using this root access, attackers can establish a connection with the wearable devices and pass certain commands to recognize the users' face, record the footage containing video, and voice recording of the user.


\vspace{-0.3cm}  
\subsection{Summary of the Existing Attacks}

\begin{figure*}[t!]
\vspace{-0.1in}
  \centering
    \subfloat[]{\includegraphics[width=0.29\textwidth]{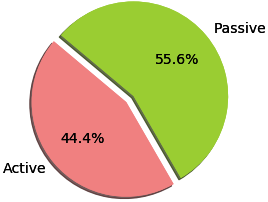}\label{approach}}
    \subfloat[]{\includegraphics[width=0.27\textwidth]{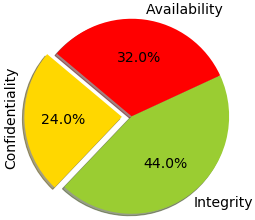}\label{security}}
    \subfloat[]{\includegraphics[width=0.29\textwidth]{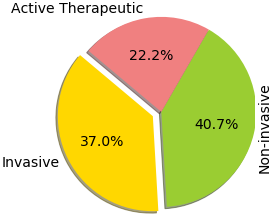}\label{device}}\\
    \subfloat[]{\includegraphics[width=0.29\textwidth]{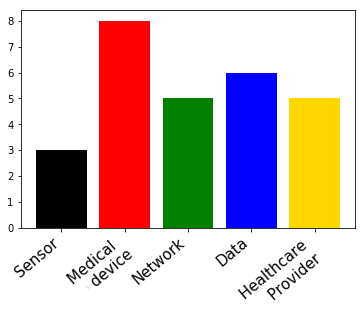}\label{component}}
    \subfloat[]{\includegraphics[width=0.29\textwidth]{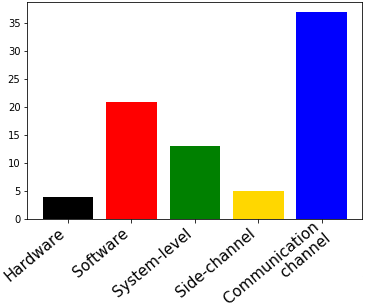}\label{threat}}
    \vspace{-0.1in}
      \caption{Summary of attacks based on (a) attack approach, (b) impacted security, (c) targeted medical devices, (d) targeted components, (e) types of attacks.}
     \label{fig:summary}
      \vspace{-0.2in}
\end{figure*}

We categorize 80 reported attacks on healthcare systems reported by the research community and developers in five categories. Additionally, we explain the attack methods and discuss the impacts of these existing attacks based on different metrics (i.e., attack approach, impacted security, targeted medical devices, targeted components, and types of attacks.). Here, we summarize some interesting findings about these attacks with the help of Figure~\ref{approach}:

\vspace{4pt}
        \noindent\textbf{Attack approach:} Based on the attack approach in a healthcare system, the attacks can be categorized as active or passive. Active attacks (e.g., DoS attack, MITM attack, etc.) try to change the healthcare system resources or affect the system's operations while passive attacks (e.g., eavesdropping, weak authentication scheme exploitations, etc.) read or make use of the information from the healthcare system resource. From Figure~\ref{approach}, one can observe that passive attacks have been reported more (55.6\%) in healthcare than active attacks (44.4\%).

\vspace{4pt}
        \noindent\textbf{Impacted security:} After a successful exploitation, confidentiality (C), integrity (I), and/or availability (A) of a healthcare system are impacted by the existing attacks. Active attacks (e.g., malware, ransomware) mostly affect the integrity and availability of a healthcare system. On the contrary, passive attacks (e.g., eavesdropping, MIMO) jeopardize the confidentiality of a healthcare system. As Figure~\ref{security} shows, the integrity of the systems is impacted the most (44.0\%) due to reported attacks in healthcare.

\vspace{4pt}
        \noindent\textbf{Targeted medical devices:} Based on the existing attacks in a healthcare system, as Figure~\ref{device} presents, \textit{non-invasive} medical devices are the most targeted by the attackers. Most of the \textit{non-invasive} devices (e.g., smartwatches, BCI devices, etc.) do not use encryption and authentication mechanisms during their communication with the programmer rather perform clear-text data transmission. As a consequence, an attacker can perform active attacks (e.g., DoS attack, replay attack, etc.) and passive attacks (e.g., eavesdropping, impersonation, etc.) on these devices. Attacks on \textit{active therapeutic} devices are much lower compared to the attacks on \textit{invasive} devices.
        
\vspace{4pt}
        \noindent\textbf{Targeted components:} Among the five components in a healthcare system (explained in Section~\ref{background}), the medical devices and data are the most affected due to attacks (Figure~\ref{component}). As the patients' data and information produced by healthcare devices can be used for blackmail and extortion~\cite{black}, attackers target medical devices and data the most. Besides these components, network and healthcare provider attacks are also high in numbers. Sensor-level attacks are the minimum in number compared to the other targeted components.

\vspace{4pt}
        \noindent\textbf{Types of attacks:} Based on the current attacks, as Figure~\ref{threat} presents, communication channel attacks are the most common attack on healthcare systems. Healthcare devices use various wireless communication protocols (e.g., Wi-Fi, Bluetooth, Zigbee, etc.) to check the patient's status remotely, which at the same time makes them vulnerable to the attacks. Software attacks are emerging as more medical devices are getting smart and support third-party applications. Hardware attacks occur because third-party vendors develop most of the ICs. Although small in number, hardware attacks are still a concern for the patients as it is increasing day by day.
        

\section{Current security and privacy solutions}\label{solutions}


In this section, we discuss the security measures on healthcare systems that have been proposed by the research communities and developers to defend against the attacks presented in Section \ref{threats}. Most of the security measures directly address the trade-offs in healthcare devices, while others propose several countermeasures against specific types of attacks. Several of these solutions are widely used or considered as standards. In most cases, the primary concern of a security measure is how to deal with the emergency when there is a risk of a patient's safety. 
In the following subsections, we discuss different existing security measures for healthcare systems by categorizing in five broad categories. A summary of these security and privacy solutions for healthcare systems is presented in Tables~\ref{table:Solution_list1} and~\ref{table:Solution_list2}.


\begin{table*}[t!]
\caption{List of the existing security and privacy solutions for healthcare devices and applications.}
\vspace{-0.1cm}
\label{table:Solution_list1}
\fontsize{16}{16}\selectfont
\resizebox{1\textwidth}{!}{
\begin{tabular}{cccp{13cm}p{12cm}c}
\toprule
\multicolumn{1}{c}{\begin{tabular}[c]{@{}c@{}}Solution \\ Type\end{tabular}} & \multicolumn{1}{c}{\begin{tabular}[c]{@{}c@{}}Defense\\ Mechanisms\end{tabular}} & \multicolumn{1}{c}{\begin{tabular}[c]{@{}c@{}}Attack \\ Type\end{tabular}} & \multicolumn{1}{c}{\begin{tabular}[c]{@{}c@{}}Summary\end{tabular}} & \multicolumn{1}{c}{\begin{tabular}[c]{@{}c@{}}Limitations\end{tabular}} & \multicolumn{1}{c}{\begin{tabular}[c]{@{}c@{}}Ref.\end{tabular}} \\ \hline
	\midrule
\multirow{54}{*}{\begin{tabular}[c]{@{}c@{}}Side-channel \\analysis\end{tabular}} & \multirow{4}{*}{\begin{tabular}[c]{@{}c@{}}Current\\ analysis\end{tabular}} & \multirow{4}{*}{\begin{tabular}[c]{@{}c@{}}Hardware\end{tabular}}
& 
\begin{itemize}
    \item Static CMOS gates are subject to leakage current in the idle mode.
    \item Needs to measure the current from multiple power pins.
\end{itemize}
&
\begin{itemize}
    \item Unable to detect small HTs because of power and timing variations.
\end{itemize}
& \multirow{4}{*}{\begin{tabular}[c]{@{}c@{}}\cite{bhunia2014hardware, aarestad2010detecting, wei2013undetectable}\end{tabular}} \\ \cline{2-6}
& \multirow{4}{*}{\begin{tabular}[c]{@{}c@{}}Delay variation \\and characterization \\techniques \end{tabular}} & \multirow{4}{*}{\begin{tabular}[c]{@{}c@{}}Hardware\end{tabular}}
	& 
	\begin{itemize}
	    \item High-precision, low-overhead embedded test structure (REBEL used to detect anomalies in Hardware Trojans (HTs) .
	    \item Capable to deliver high-resolution measurement of path-delays.
	    \item Linear regression used to classify delay behavior
	     \item Apply on a large number of unobservable internal combinatorial register-to-register paths.
	\end{itemize}
	 & 
	 \begin{itemize}
	     \item Timing-based HT a backtracking-based algorithm used to identify reconvergent paths in the circuit that unable to catch the delay variations caused by HT. 
	     \item Additional power and area overheads may not acceptable for medical devices.
	 \end{itemize}
	 &\multirow{6}{*}{\begin{tabular}[c]{@{}c@{}}\cite{lamech2012trojan, wei2012hardware}\\ \cite{li2008speed}\end{tabular}} \\ \cline{2-6}
	 
	& \multirow{4}{*}{\begin{tabular}[c]{@{}c@{}}Battery-constraint \\mitigation\end{tabular}} & \multirow{4}{*}{\begin{tabular}[c]{@{}c@{}}Communication\\ channel\end{tabular}}
	& 
	\begin{itemize}
	    \item BAN protocols can be used to mitigate this problem. 
	    \item The node will not wake up without any messages that are outside of negotiated time intervals.
	    \item Data-independent power consumption as circuit level solutions can be used here.
	    \item A security protocol called IMDfence can be used here.
	\end{itemize}
	 & 
	 \begin{itemize}
	     \item Jamming based protection is provided by many external security devices, but it is not always effective because it can cause battery depletion.
	 \end{itemize}
	 & \multirow{4}{*}{\begin{tabular}[c]{@{}c@{}}\cite{halperin2008pacemakers},\\\cite{kwak2010overview, tiri2002dynamic, tiri2004charge, siddiqi2020imdfence, siddiqi2019towards}\end{tabular}} \\ \cline{2-6}
	& \multirow{4}{*}{\begin{tabular}[c]{@{}c@{}}Power Consumption \\Analysis\end{tabular}} & \multirow{4}{*}{\begin{tabular}[c]{@{}c@{}}Software\end{tabular}}
	& 
	\begin{itemize}
	    \item Measures the power consumption of traditional programs and known malware. 
	    \item Compares them with the power signatures to find any anomaly in the power consumption. 
	    \item Any aberration in power consumption can be detected as anomaly behavior in case of medical devices.
	\end{itemize}
	 & 
	 \begin{itemize}
	     \item No consistent base set of known-good behaviors on a PC.
	     \item A system like WattsUpDoc would likely to raise false alarms because of an inconsistent or inaccurate internal model.
	 \end{itemize} & \multirow{4}{*}{\begin{tabular}[c]{@{}c@{}}\cite{clark2013wattsupdoc}\end{tabular}} \\ \cline{2-6}
	 
	& \multirow{4}{*}{\begin{tabular}[c]{@{}c@{}}Shielding and \\Filtering\end{tabular}} & \multirow{4}{*}{\begin{tabular}[c]{@{}c@{}}Side-channel\end{tabular}}
	& 
	\begin{itemize}
	    \item The exterior of a healthcare device was covered with with a conducting surface.
	    \item Attackers were forced to transmit $10^4$ times more powerful signal to have the same effect as before.
	    \item Faraday cage is another countermeasure to block EM radiation.
	\end{itemize}
	 & 
	 \begin{itemize}
	     \item Very difficult to defeat electromagnetic analysis, except if the circuit and its countermeasures are overlapped.
	 \end{itemize}
	  & \multirow{4}{*}{\begin{tabular}[c]{@{}c@{}}\cite{kune2013ghost}, \cite{quisquater2001electromagnetic}\end{tabular}}  \\ \cline{2-6}
	& \multirow{4}{*}{\begin{tabular}[c]{@{}c@{}}Masking\end{tabular}} & \multirow{4}{*}{\begin{tabular}[c]{@{}c@{}}Side-channel\end{tabular}}
	& 
	\begin{itemize}
	    \item Intermediate values of a cryptographic computation are randomized by masking.
	    \item A key masking method can be used against DPA. 
	    \item . A band-pass filter or a current-flattening circuit can be added to the cryptosystem to suppress information leakage.
	\end{itemize}
	 & 
	 \begin{itemize}
	     \item Very difficult to defeat electromagnetic analysis because energy overhead is a big concern for healthcare devices.
	 \end{itemize}
	  & \multirow{4}{*}{\begin{tabular}[c]{@{}c@{}}\cite{ratanpal2004chip, hasan2001power,muresan2008protection, liu2010low}\end{tabular}}  \\ \cline{2-6}
	  	 \hline
\multirow{20}{*}{\begin{tabular}[c]{@{}c@{}}Hardware-centric \\solution\end{tabular}} & \multirow{4}{*}{\begin{tabular}[c]{@{}c@{}}Hardware Trojan \\triggering and \\PUF\end{tabular}} & \multirow{4}{*}{\begin{tabular}[c]{@{}c@{}}Hardware, \\ Communication \\ channel\end{tabular}}
& 
	\begin{itemize}
	    \item Testing the design right after the chip fabrication step.
	    \item Detect HTs with significant footprints like a golden die method.
	    \item PUF method derives a secret from the physical characteristics of the IC.
	\end{itemize}
	 & 
	 \begin{itemize}
	     \item Cleverly inserted HTs may not be easily triggered. 
	     \item Does not guarantee the correctness of the design at run-time.
	     \item Offline full functionality test is inefficient and time-consuming. 
	 \end{itemize}
	 & \multirow{4}{*}{\begin{tabular}[c]{@{}c@{}}\cite{wu2016tpad,francq2015introduction, herder2014physical}\end{tabular}} \\ \cline{2-6}
	& \multirow{4}{*}{\begin{tabular}[c]{@{}c@{}}Physical \\Separation\end{tabular}} & \multirow{4}{*}{\begin{tabular}[c]{@{}c@{}}Software \end{tabular}}
	&
	\begin{itemize}
	    \item Executes critical security applications on isolated hardware that is free of observation and interference through direct physical access.
	\end{itemize}
	 &
	 \begin{itemize}
	     \item Does not hide the data processing which may reveal the patient information. 
	     \item Vulnerable to DoS attack. 
	 \end{itemize}
	 & \multirow{4}{*}{\begin{tabular}[c]{@{}c@{}}\cite{aaraj2008analysis, sorber2012plug}\end{tabular}} \\ \cline{2-6}
	 & \multirow{4}{*}{\begin{tabular}[c]{@{}c@{}}Online HT \\ detection method\end{tabular}} & \multirow{4}{*}{\begin{tabular}[c]{@{}c@{}}Hardware\end{tabular}}
    &
    \begin{itemize}
        \item Identify HTs by checking underlying hardware functionality.
        \item Architecture is divided into two-chip generating signatures deep in the hardware and later check it during data processing and transmission.
    \end{itemize}
     &
     \begin{itemize}
         \item Relied on the digital logic modules which minimally impacts the performance of the system.
     \end{itemize}
      &\multirow{4}{*}{\begin{tabular}[c]{@{}c@{}}\cite{wehbe2018securing}\end{tabular}} \\ \cline{2-6}
      \hline
	\multirow{13}{*}{\begin{tabular}[c]{@{}c@{}}Software-centric \\solution\end{tabular}} & \multirow{4}{*}{\begin{tabular}[c]{@{}c@{}}Secure execution \\environment\end{tabular}} & \multirow{4}{*}{\begin{tabular}[c]{@{}c@{}}Software \\ System-level\end{tabular}}
& 
    \begin{itemize}
        \item Secure virtual machines (VM) provides a secure network interface, secure storage, secure execution environment.
    \end{itemize}
     &
     \begin{itemize}
         \item  Management environment can still be a compromised operating system. 
         \item Performance penalties are based on execution-specific domain operations as well as several benchmarks.
     \end{itemize}
     & \multirow{4}{*}{\begin{tabular}[c]{@{}c@{}}\cite{li2010secure}\end{tabular}} \\ \cline{2-6}
	& \multirow{4}{*}{\begin{tabular}[c]{@{}c@{}}Static Analysis\end{tabular}} & \multirow{4}{*}{\begin{tabular}[c]{@{}c@{}}Software\end{tabular}}
	& 
	\begin{itemize}
	    \item Provides almost complete coverage of the code and helps to detect potentially fatal errors.
	    \item May not easily be detected through conventional testing methods, e.g. CodeSonar.
	\end{itemize}
	 & 
	 \begin{itemize}
	     \item It identifies many bugs in the software but static analysis tools are not a replacement for testing.
	 \end{itemize}
	 & \multirow{4}{*}{\begin{tabular}[c]{@{}c@{}}\cite{jetley2008static}\end{tabular}} \\ \cline{2-6}
\end{tabular}}
    \vspace{-0.5cm}
\end{table*}
    

\begin{table*}[t!]
\caption{List of the existing security and privacy solutions for healthcare devices and applications (Continued).}
\vspace{-0.2cm}
\label{table:Solution_list2}
\fontsize{15.5}{15.5}\selectfont
\resizebox{1\textwidth}{!}{
\begin{tabular}{cccp{13cm}p{12cm}c}
\toprule
\multicolumn{1}{c}{\begin{tabular}[c]{@{}c@{}}Solution \\ Type\end{tabular}} & \multicolumn{1}{c}{\begin{tabular}[c]{@{}c@{}}Defense\\ Mechanisms\end{tabular}} & \multicolumn{1}{c}{\begin{tabular}[c]{@{}c@{}}Attack \\ Type\end{tabular}} & \multicolumn{1}{c}{\begin{tabular}[c]{@{}c@{}}Summary\end{tabular}} & \multicolumn{1}{c}{\begin{tabular}[c]{@{}c@{}}Limitations\end{tabular}} & \multicolumn{1}{c}{\begin{tabular}[c]{@{}c@{}}Ref.\end{tabular}} \\ \hline
	\midrule
	\multirow{18}{*}{\begin{tabular}[c]{@{}c@{}}Software-centric \\solution\end{tabular}} & \multirow{4}{*}{\begin{tabular}[c]{@{}c@{}}Run-time \\Analysis\end{tabular}} & \multirow{4}{*}{\begin{tabular}[c]{@{}c@{}}Software\end{tabular}}
	& 
	\begin{itemize}
	    \item A dynamic binary instrumentation-based malware detection framework.
	    \item Can trace the untrusted program during execution time in a virtualized testing environment.
	    \item With a large number of input values and with extensive security policies can detect malware behavior.
	\end{itemize}
	 & 
	 \begin{itemize}
	     \item Depends on the accuracy of the security policies used.
	     \item Along with the number of observed paths, in particular, observed malicious paths, in the Testing environment.
	 \end{itemize}
	  & \multirow{4}{*}{\begin{tabular}[c]{@{}c@{}}\cite{aaraj2008dynamic}\end{tabular}} \\ \cline{2-6}
	& \multirow{4}{*}{\begin{tabular}[c]{@{}c@{}}Formal \\Verification\end{tabular}} & \multirow{4}{*}{\begin{tabular}[c]{@{}c@{}}Software\end{tabular}}
	&
	\begin{itemize}
	    \item Functional specification of the medical device are expressed in input/output sequences
	    \item Then translated into assert verifiable property so that medical device software accept it. 
	    \item A model checker is used to fed transformed code and valid assertions. 
	\end{itemize}
	 & 
	 \begin{itemize}
	     \item Current software verification tools have written in a high-level programming language.
	     \item Not suitable for highly platform specific and low-level programs.
	     \item System-level properties need to be verified with the real world medical device interfaces.
	 \end{itemize}
	  & \multirow{4}{*}{\begin{tabular}[c]{@{}c@{}}\cite{li2013improving, cordeiro2009semiformal, jetley2006formal}\end{tabular}} \\ \cline{2-6}
	\hline
\multirow{25}{*}{\begin{tabular}[c]{@{}c@{}}Trust-management \\framework\end{tabular}}	& \multirow{4}{*}{\begin{tabular}[c]{@{}c@{}}Biometrics\end{tabular}} & \multirow{4}{*}{\begin{tabular}[c]{@{}c@{}}Communication\\ channel\end{tabular}}
	& 
	\begin{itemize}
	    \item Electrocardiography signal asserts the time between heartbeats, or interpulse interval (IPI). 
	    \item Creates a high level of randomness and can be measured from anywhere on the body. 
	    \item Temporal and morphological alterations of ECG measurements could be detected using Arterial Blood Pressure (ABP) signal.
	\end{itemize}
	 & 
	 \begin{itemize}
	     \item If any physiological signals stay within the human body is incorrect, both the security and privacy of schemes may be affected.
	 \end{itemize}
	& \multirow{4}{*}{\begin{tabular}[c]{@{}c@{}}\cite{poon2006novel, cornelius2012wears} \\ \cite{hu2013opfka, venkatasubramanian2010pska, chang2012body, rostami2013heart, jurik2011securing, cherukuri2003biosec, shafiul2020prediction, chizari2019extracting, ashrafi2015modified, belkhouja2019biometric, cai2019data, cai2016detecting}\end{tabular}} \\ \cline{2-6}
	& \multirow{4}{*}{\begin{tabular}[c]{@{}c@{}}Out-of-Band \\(OOB) Authentication\end{tabular}} & \multirow{4}{*}{\begin{tabular}[c]{@{}c@{}}Communication\\ channel\end{tabular}}
	& 
	\begin{itemize}
	    \item Use audio, visual or haptic channels for authentication. 
	    \item Visual OOB authentication, e.g. ultra-violet or visible tattoos to record permanent implantable medical devices keys.
	\end{itemize}
	 & 
	 \begin{itemize}
	     \item Auxiliary channels can be breached by close range eavesdropping. 
	     \item Visual OOB suitable for the emergency but will be a problem for key revocation and usability concern.
	 \end{itemize}
	 & \multirow{4}{*}{\begin{tabular}[c]{@{}c@{}}\cite{halperin2008pacemakers}, \\\cite{denning2010patients, li2013secure, schechter2010security, goodrich2006loud} \end{tabular}} \\ \cline{2-6}
	& \multirow{4}{*}{\begin{tabular}[c]{@{}c@{}}Close-range \\Communication\end{tabular}} & \multirow{4}{*}{\begin{tabular}[c]{@{}c@{}}Communication\\ channel\end{tabular}}
	& 
	\begin{itemize}
	    \item Near-field communication (NFC) and RFID-based channel can be utilized here. 
	    \item Distance bounding communication between the medical device and external devices. 
	    \item Access is granted only if the devices are within a safe range.
	\end{itemize}
	 & 
	 \begin{itemize}
	     \item A successful RFID eavesdropping attack is possible at a distance of a few meters with off-the-shelf antenna kits
	 \end{itemize}
	 & \multirow{4}{*}{\begin{tabular}[c]{@{}c@{}}\cite{israel2001pacemaker, freudenthal2007suitability, baldus2009human, bagade2013protect, rasmussen2009proximity, shi2013bana}, \\\cite{cremers2012distance}\end{tabular}}  \\ \cline{2-6}
	& \multirow{4}{*}{\begin{tabular}[c]{@{}c@{}}External Device\end{tabular}} & \multirow{4}{*}{\begin{tabular}[c]{@{}c@{}}Communication\\ channel\end{tabular}}
	&
	\begin{itemize}
	    \item Ensure radio security without any modification to the IWMD itself. 
	    \item Patients ECG signals used in IMDGuard to extract keys explicitly. 
	    \item Physical characteristics such as RSSI, time of arrival , differential time of arrival , and angle of arrival are used to detect anomalies.
	\end{itemize}
	 & 
	 \begin{itemize}
	     \item Jamming based protection not always effective.
	 \end{itemize}
	  & \multirow{4}{*}{\begin{tabular}[c]{@{}c@{}}\cite{tippenhauer2013limitations},\\\cite{denning2008absence, gollakota2011they, xu2011imdguard, zhang2013medmon}\end{tabular}} \\ \cline{2-6}
\hline
	\multirow{34}{*}{\begin{tabular}[c]{@{}c@{}}Data protection \\solution\end{tabular}} & \multirow{8}{*}{\begin{tabular}[c]{@{}c@{}}Encryption\end{tabular}} & \multirow{8}{*}{\begin{tabular}[c]{@{}c@{}}Communication\\ channel\end{tabular}}
	& 
    \begin{itemize}
        \item PRESENT and KATAN lightweight hardware-oriented block ciphers are as small as 1000-1500 gate equivalent for a 64-bit block size encryption. 
        \item Stream mode utilizes output feedback (OFB) to obtain a scalable stream cipher. 
        \item Encompression is the combination of compressive sensing, encryption and integrity checking.
        \item Symmetrical encryption algorithms are the distribution of shared key between devices.
    \end{itemize}
     & 
     \begin{itemize}
         \item Low power symmetric ciphers still may increase the energy consumption which will shorten the battery life. 
         \item If the patient is unconscious, then key distribution needs to be done without the patient's intervention for timely treatment.
     \end{itemize}
      
    & \multirow{6}{*}{\begin{tabular}[c]{@{}c@{}}\cite{zhang2014trustworthiness}, \cite{denning2010patients}, \\\cite{schechter2010security},\\\cite{de2009katan, potlapally2003analyzing, bogdanov2007present,donoho2006compressed, heron2009advanced, bertoni2013keccak, shi2013ask,ali2012zero, jana2009effectiveness,mathur2008radio, hosseini2011lightweight, rostami2013balancing, beck2011block, zhang2013energy, bu2019bulwark}\end{tabular}} \\ \cline{2-6}
    & \multirow{8}{*}{\begin{tabular}[c]{@{}c@{}}Machine Leaning\\-based approaches\end{tabular}} & \multirow{8}{*}{\begin{tabular}[c]{@{}c@{}}Communication\\ channel\end{tabular}}
	& 
    \begin{itemize}
        \item A decision tree algorithm was used to detect malicious attacks in healthcare devices. 
        \item Support Vector Machine (SVM), Trees, and Ensemble algorithms were used to authenticate users for healthcare devices.
        \item A machine learning-based security framework was proposed to detect malicious activities in a healthcare system.
    \end{itemize}
     & 
     \begin{itemize}
         \item Energy consumption is an issue for resource limited healthcare devices. 
     \end{itemize}
      
    & \multirow{6}{*}{\begin{tabular}[c]{@{}c@{}}\cite{saeedi2019machine, vhaduri2017wearable, newaz2019healthguard, hasan2019supervised, rathore2020deep}\end{tabular}} \\ \cline{2-6}
    & \multirow{8}{*}{\begin{tabular}[c]{@{}c@{}}Access control\\-mechanisms\end{tabular}} & \multirow{8}{*}{\begin{tabular}[c]{@{}c@{}}System-level\end{tabular}}
	& 
    \begin{itemize}
        \item An identity-based access control proposed to protect private health data in a cloud-assist health monitoring system.
        \item Attribute-based access policy has been employed in many research works to control access to medical data, BAN, cloud system. 
        \item Proximity-based access control is based on the distance between the programmer and the healthcare device. 
    \end{itemize}
     & 
     \begin{itemize}
         \item Most access control mechanisms focused only on the healthcare device access authentication.
     \end{itemize}
      
    & \multirow{6}{*}{\begin{tabular}[c]{@{}c@{}} \cite{li2011hijacking, halperin2008pacemakers}, \\ \cite{sun2011hcpp, lin2013cam, li2010data, li2012scalable, guan2015achieving, hei2010defending, zhang2013medmon, fu2019poks}\end{tabular}} \\ \cline{2-6}
    & \multirow{8}{*}{\begin{tabular}[c]{@{}c@{}}Blockchain-based\\approaches\end{tabular}} & \multirow{8}{*}{\begin{tabular}[c]{@{}c@{}}Communication\\ channel\end{tabular}}
	& 
    \begin{itemize}
        \item A hybrid approach that combined advantages of the private and public key, blockchain, and many other lightweight cryptographic primitives to develop a patient-centric access control for electronic medical records.
        \item Researchers introduced a reliable data communication and storage with more advanced and lightweight cryptographic techniques like the ARX encryption scheme.
    \end{itemize}
     & 
     \begin{itemize}
         \item Storage is an issue for the resource-constrained healthcare devices. 
     \end{itemize}
      
    & \multirow{6}{*}{\begin{tabular}[c]{@{}c@{}}\cite{chen2019blockchain, dwivedi2019decentralized, srivastava2019light, shahid2019sensor, srivastava2019data}\end{tabular}} \\ \cline{2-6}
	  \hline
	
\end{tabular}}
    \vspace{-0.5cm}
\end{table*}
\vspace{-0.4cm}
\subsection{Solutions based on Side-channel Analysis}
    
The side-channel analysis relies on analyzing the information gained from the physical properties of the healthcare devices (e.g., energy consumption, timing analysis, or electromagnetic emanations, etc.) and compares the information with the data generated by the normal behavior of a device to detect anomalies. Most of these solutions can be broadly divided into the following sub-categories: electric current analysis, delay variation and characterization, power consumption analysis, shielding and filtering, masking, and battery-constraint mitigation. 

\vspace{6pt}
\noindent\textbf{Electric current analysis:} The electric current analysis measures the current consumption in a device, allowing experts to detect anomalies in devices at the hardware level. Bhunia et al. monitored current leakage from static CMOS gates to identify trojan circuits in a healthcare device~\cite{bhunia2014hardware}. As current leakage always remains the same for CMOS gate, the difference in current consumption can distinguish hardware trojan from the base circuits. However, in the case of a large circuit with a high number of gates and fewer trojan, the current analysis fails due to an insignificant change in current readings and difficulty of performing extensive testing. Aarestad et al. proposed current analysis in multiple pins instead of a single point to increase the sensitivity and reduce the problem of detecting a few gates in a fraction of the total gates in the IC~\cite{aarestad2010detecting}. However, this technique cannot detect HTs that are small in size due to the power and timing variations that HT can cause \cite{wei2013undetectable}.
        
\vspace{6pt}
\noindent\textbf{Delay variation and characterization:} This 
HT detection method works by measuring and detecting small systematic changes in path delays introduced by capacitive loading effects or series inserted gates of HTs. A high-precision, low-overhead embedded test structure (ETS) called \textit{REBEL} was proposed to detect delay anomalies in HTs~\cite{lamech2012trojan}. 
\textit{REBEL} was capable of delivering high-resolution measurements of path delays and able to identify a wide range of delay anomalies introduced by HTs. It provides significant benefits over other traditional delay testing methods as the digital snap-shot captured by REBEL allows glitches to be detected and can potentially speed up the path delay measurements using a small number of repeated application of the test pattern. The detection sensitivity of REBEL was checked by varying the analog control voltage on each trojan emulation circuits one at a time and classifying the result using regression analysis. A backtracking-based algorithm was proposed in \cite{wei2012hardware} to identify the reconvergent locations in the circuits where the delay variations caused by HT is not observable.
        
A new method for IC authentication and hardware trojan horse (HTH) detection is delay characterization technique introduced in \cite{li2008speed}. Such a technique measures the combination of an arbitrarily large number of register-to-register paths delays internal to the functional portion of the 
IC. This technique is originally developed to apply on a large number of unobservable internal combinatorial register-to-register paths to get accurate, precise data about path delays. Moreover, this technique can also be used for HTH detection by extracting the non-functional path delay characteristics to detect malicious circuit alterations. The delay measurement technique does not affect the circuit functionality and allows to monitor delay characteristics at run-time. However, this technique needs to assess the authentication approach across a large number of physical ICs, which introduces time latency and resource overhead. As healthcare devices are power-constraining devices, additional power, and area overheads may not be accepted.

\vspace{6pt}
\noindent\textbf{Power consumption analysis:} Malware detection in healthcare devices can be performed by power consumption analysis, which compares the power consumption of traditional programs and known malware to find anomaly in the system. 
Clark et al. proposed a behavior monitoring system, \textit{WattsUpDoc}, which uses supervised learning to classify normal and abnormal power consumption in the replacement of a pre-constructed power consumption model and detect anomaly behavior in healthcare devices~\cite{clark2013wattsupdoc}. The key idea is to high sampling rate in power consumption tracing and high-accuracy power measurement to detect malware in the healthcare devices. However, \textit{WattsUpDoc} raises false alarms in case of an inconsistent base set of known-behavior. Also, achieving high accuracy power measurement in resource-constrained medical devices is difficult, which results in a low accuracy rate in detecting malware.
        
\vspace{6pt}
\noindent\textbf{Battery-constraint mitigation:} As healthcare devices are resource-constraint devices, battery depletion/drainage attacks can cause severe obstruction in the normal operation of the devices. However, a defense mechanism against the battery drainage attack needs to be power-efficient, or else the defense mechanism itself may consume more power than the attack. Researchers proposed zero-power defenses for ICDs where RF energy harvested from external sources are used for notification, authentication, and key exchange~\cite{halperin2008pacemakers}. Also, BAN protocols can be used to mitigate this problem. For instance, the IEEE 802.15.6 BAN standard allows a node and hub to negotiate their communication intervals by encoding them in authenticated messages. Accordingly, the node will not wake up without any messages that are outside of negotiated time intervals to save power in the devices \cite{kwak2010overview}. Tiri et al. proposed novel logic styles with data-independent power consumption as circuit-level solutions against battery drainage attacks to reduce the dependence of power dissipation on input patterns~\cite{tiri2002dynamic, tiri2004charge}. 
In recent work, Siddiqi et al. proposed an adaptive zero-power defense solution using a radio frequency power transfer mechanism based on energy harvesting against battery-depletion attacks~\cite{siddiqi2019towards}.

\vspace{6pt}
\noindent\textbf{Shielding and filtering:} Shielding and filtering are commonly used to defend against EMI attacks. Kune et al. showed that covering the exterior of a healthcare device with a conducting surface can force the attacker to transmit $10^4$ times more powerful signal to have the same effect in an EMI attack, which can be differentiated from legitimate signals easily~\cite{kune2013ghost}. Faraday cage is another countermeasure to block EM radiation and to reduce the EM radiation signature \cite{quisquater2001electromagnetic}. However, it is hard to defeat electromagnetic analysis, except if the circuit and its countermeasures overlapped.

\vspace{6pt}
\noindent\textbf{Masking:} Masking is a technique of hiding original data with modified content. Intermediate values of a cryptographic computation are randomized by masking, which avoids dependencies between these values and the power consumption applied in algorithmic level. Moreover, it does not rely on the power consumption characteristics of the medical device. Researchers proposed a key masking method as a software solution against DPA attacks \cite{hasan2001power}. Although this method attempts to randomize the secret key before each execution of the scalar multiplication, power overhead is a concern here for healthcare devices. A band-pass filter \cite{ratanpal2004chip} or a current-flattening circuit \cite{muresan2008protection} can be added to the cryptosystem to suppress information leakage through the current supply pin. An internally generated random mask based on ring oscillators was proposed in \cite{liu2010low} to change the power consumption dynamically.

\vspace{-0.4cm}    
\subsection{Hardware-centric Solutions}
Hardware-centric solutions are designed to protect healthcare devices from hardware-level attacks like hardware trojans. The solutions are broadly divided into several categories, including hardware trojan triggering and physical unclonable function (PUF), physical separation, and online HT detection method.
    
\vspace{6pt}
\noindent\textbf{Hardware trojan triggering and PUF:} 
This type of HT detection depends on testing the design after the chip fabrication. Wu et al. proposed \textit{Golden Die} Method to detect HTs with significant footprints differentiated from the base standard~\cite{wu2016tpad}. However, cleverly inserted HTs may not be easily triggered by this approach as the testing mechanism may not know the presence of HT and its location in the chip. Francq et al. presented a functional verification method that depends on checking the functionality of the hardware by monitoring the output and checking for expected behavior~\cite{francq2015introduction}. Compare to the aforementioned methods, the PUF method is useful for healthcare applications as it derives a secret from the physical characteristics of the IC instead of storing secrets in digital memory \cite{herder2014physical}. 
    
\vspace{6pt}
\noindent\textbf{Physical separation:} This is a hardware-centric design solution where security applications run on separate hardware from the device's main architecture. The trusted platform module (TPM) is proposed as a physical separation method where the cryptographic keys for a specific host computer are stored in a separate module to use in IWMDs. As separate module introduces overhead, a software-based TPM can be used to increase the effectiveness in healthcare devices~\cite{aaraj2008analysis}. Sorber et al. proposed \textit{Plug-n-Trust}, which is a MicroSD card that provides a trusted computing platform on a smartphone~\cite{sorber2012plug}. \textit{Plug-n-Trust} encrypts all the medical data transmitted from the devices and can only be decrypted in the card for further analysis in verified API. However, it does not hide data processing, which may reveal the patient's information and can be vulnerable to DoS attacks.

\vspace{6pt}
\noindent\textbf{Online HT detection method:} Online HT detection method refers to identify HTs at run-time. Wehbe et al. proposed an online method to identify HTs by checking underlying hardware functionality at run-time \cite{wehbe2018securing}. Here, researchers divided the whole architecture in two-chip, generating signatures deep in the hardware and later checked it during data processing and transmission. This technique relies on the digital logic modules that minimally impacts the performance of the system.
        

\vspace{-0.2cm}    
\subsection{Software-centric Solutions}
    
Software-centric solutions are designed to protect healthcare devices from software-level and system-level attacks (e.g., malware, ransomware, weak authentication scheme exploitations, counterfeit firmware, etc.). The software-centric solutions can be categorized as follows: secure execution environment, static analysis, run-time monitoring, and formal verification. 
    
\vspace{6pt}
\noindent\textbf{Secure execution environment:} A secure execution environment is a safe execution space for executing the code that ensures security to the code and loaded data. For ensuring the safety of healthcare applications, one needs to run these applications on a secure execution environment to protect from a compromised OS. In this sense, secure virtual machines (VM) can be a viable solution that provides a network interface, storage, and execution environment. The main advantage of using a secure VM is that only security-critical healthcare applications will be running on the VM, making them isolated from other applications that may be compromised. However, the management environment can still be a compromised OS \cite{li2010secure}. 
        
\vspace{6pt}
\noindent\textbf{Static analysis:} Although a secure execution environment can protect healthcare applications from a compromised system, it can not defend them if the healthcare application itself is a malware. Static analysis techniques can be used to characterize the execution behavior of a program and find program flaws by analyzing the source code. By using symbolic execution techniques to explore the execution paths of the software, static analysis detects potentially fatal errors that may not be easily detected through conventional testing methods. Jetley et al. proposed \textit{CodeSonar}, a static analysis tool to automatically detect buffer overrun, initialized variables and null pointer dereference in the healthcare apps \cite{jetley2008static}. However, static analysis depends on the availability of the source code, which is not openly available for healthcare applications. 
        
\vspace{6pt}
\noindent\textbf{Run-time analysis:} Run-time analysis can be used to detect unintended behavior of a healthcare app at run-time. Aaraj et al. proposed a dynamic binary instrumentation-based (DBI) malware detection framework that observes the execution of unknown programs, models safe/unsafe behavior with respect to specified security policies, and ensures that the program does not deviate from safe behavior \cite{aaraj2008dynamic}. This framework uses two virtual execution environments, testing and real environment. In the testing environment, DBI collects specific information in the form of execution traces to construct a hybrid model for representing dynamic control and data invariants. 
This hybrid model along with a behavioral model generated from security polices is used to define the malware behavior in DBI framework. In the test environment, a program is analyzed based on the generated model in test environment to detect malicious behavior of a program in healthcare devices. An unknown program is only moved into the real environment to monitor its execution if the behavior pattern is matched with the allowed model generated in test environment. 
        
\vspace{6pt}
\noindent\textbf{Formal verification:} Formal verification methods are proposed in several prior works to develop reliable medical device systems \cite{li2013improving, cordeiro2009semiformal, jetley2006formal}. Formal methods are well-formed statements in mathematical logic, and formal verification provides strict deduction of that logic. As a result, the entire state space of a system can be examined to establish a security property for all possible input. Formal verification can be used to verify whether medical devices are free from vulnerabilities or not. An example of a formal verification approach is described in \cite{li2013improving}, where medical device software is the first subject to the source transformation to address the semantic gaps. Properties based on the functional specification of the medical device are expressed in input/output sequences and then translated into a verifiable property to make it acceptable for the medical device software. After that, a model checker is used to feed transformed code, and valid assertions, which verifies the code against the statements and reports whether the code is acceptable or an anomaly is detected. Current software verification tools are written in a high-level programming language, which is not suitable for platform-specific and low-level programs of medical devices. Researchers proposed a semiformal verification approach that was a combination of dynamic and static verification to cover the state space exhaustively \cite{cordeiro2009semiformal}. A pre and post-market analysis were carried out based on formal verification techniques to support the process of reviewing healthcare software in~\cite{jetley2006formal}.
        
        
\vspace{-0.3cm}    
\subsection{Trust Management Framework for Communication Channel Attacks}
    
    
   A trust management framework focuses on securing information flow and communication among healthcare devices by certified software and application in the system. This framework can be categorized into biometrics, out-of-band authentication, close-range communication, and external devices.

\vspace{6pt}
\noindent\textbf{Biometrics:} Biometric properties such as fingerprint, EEG, heart rate, blood glucose, etc. can be used to authenticate a healthcare device and establish trust between two communicating devices worn in the same body~\cite{hu2013opfka, venkatasubramanian2010pska, chang2012body, rostami2013heart, jurik2011securing, cherukuri2003biosec, chizari2019extracting, belkhouja2019biometric}. Poon et al. presented an ECG signal based trust management system, which asserts that the time between heartbeats or interpulse interval (IPI) can create a high level of randomness and can be used to generate a secret key for communication~\cite{poon2006novel}. Cai et al. proposed a user-specific supervised learning model to detect temporal and morphological alterations of ECG measurements using arterial blood pressure (ABP) signal~\cite{cai2019data, cai2016detecting}. As ECG and ABP signals both measure the cardiac process, different physiological signals generated by the same underlying physiological process are inherently correlated. Any unilateral change in the ECG signal without a corresponding change in the ABP signal can be detected by the proposed model. A wearable sensor is used in \cite{cornelius2012wears} to authenticate users passively with high accuracy (>90\%) by measuring their bio-impedance to alternating current of different frequencies. However, specific physiological signals within the human body can be incorrect, which may affect both the security and privacy of healthcare devices. 

    
\vspace{6pt}
\noindent\textbf{Out-of-Band (OOB) authentication:} An auxiliary channel or out-of-band (OOB) communication uses audio, visual, or haptic channels for authentication that are outside the established data communication channel \cite{denning2010patients, li2013secure, schechter2010security, goodrich2006loud}. 
Halperin et al. proposed a low-frequency audio channel that enables medical devices (e.g., implantable medical devices) to use a zero-power radio-frequency identification (RFID) for generating and transmitting a key over the audio channel. Denning et al. proposed visual OOB authentication (e.g., ultra-violet or visible tattoos) to record permanent IMD keys, where the keys are only visible under UV (black) lights~\cite{denning2010patients}. Though this mechanism is suitable for the emergency situation, it will be a problem for key revocation and usability. Also, Li et al. \cite{li2013secure} proposed a mechanism where the users are required to inspect simultaneous LED blinking visually to achieve authentication in BANs. However, this is not appropriate for emergency situations if the patient is unconscious, which makes its application limited.
    
\vspace{6pt}
\noindent\textbf{Close-range communication:} To prevent unauthorized access, restricting the communication range is an intuitive way to avoid radio attacks. If a healthcare system uses close-range communication, the attacker has to come within the range to perform radio attacks which increases the chance of detecting the attack/attacker. There are several close-ranged communication (e.g., near-field communication, RFID-based channel,  near-field identification, etc.) proposed in prior works to secure the communication in a healthcare system~\cite{israel2001pacemaker, freudenthal2007suitability}. One recent close-ranged communication for securing healthcare devices is body-coupled communication (BCC) proposed in~\cite{baldus2009human}. BCC uses the human body as a signal propagation medium, which utilizes two different mechanisms: the transmission line approach and the capacitive approach. The transmission line approach uses the human body as a transmission line where electrodes are directly attached to the human body for directly transmitting the electrical signals. The capacitive approach uses the human body as a floating conductor, whose electric potential is changed with the electric field generated by the transmitter. However, the idea of using BCC is not ideal as physiological signals can be read during physical contact like handshake \cite{bagade2013protect}. An alternative solution of short-range communication can be distance bounding communication between the medical device and external devices. Here, access to external devices is granted if the devices are within a safe range. The distance can be measured in various ways, including limiting the response time to a verification request, Ultrasonic waves, received signal strength indicator, etc. \cite{cremers2012distance, rasmussen2009proximity, shi2013bana}. 
    
\vspace{6pt}
\noindent\textbf{External devices:} For enhancing the security of existing medical devices, a significant modification is required in hardware and software, which may lead to unintended changes in their behavior. To address this problem, recent studies suggest using external devices to ensure communication medium-security without any modification to the existing healthcare devices. An external device can be used as an authentication module to verify service requests from the external user, which can save the battery life of the main healthcare devices. Denning et al. proposed \textit{Cloaker }, a wearable device to block access requests from all external programmers at run-time~\cite{denning2008absence}. \textit{Cloaker} allows access to only preauthorized programmers in the normal mode while any programmer, even unauthorized one, can access the device in the emergency mode. 
    
A group of researchers proposed \textit{Shield}, which is a personal base station placed in between the IWMD and the external programmer in\cite{gollakota2011they}. \textit{Shield} works as a relay which only allows communication from legitimate programmer while jamming all other direct communication to IWMDs. 
Furthermore, \textit{Shield} provides an encryption scheme to encrypt and decrypt sensitive information shared between the programmer and the IWMD. 
As the communication is encrypted and unauthorized communication is jammed, the confidentiality of medical device messages is ensured by \textit{Shield}.
        
Xu et al. proposed \textit{IMDGuard}, an external wearable device designed for ICD to coordinate interactions between ICD and other external programmers~\cite{xu2011imdguard}. \textit{IMDGuard} uses patients' ECG signals to generate keys to share between the ICD and programmer, upon their first connection. Any other external programmers need to be verified by the IMDGuard before they can communicate with the ICD. In a recent work, Zhang et al. introduced \textit{Medmon}, an external device that detects abnormal communication to/from the IWMDs~\cite{zhang2013medmon}. \textit{Medmon} uses different physical characteristics (i.e., received signal strength indicator (RSSI), time of arrival, differential time of arrival) to detect signal anomalies in transmission and alerts users regarding the attack. \textit{Medmon} also captures behavioral abnormalities such as vital signs or commands that lie outside the historical records of the patient. However, jamming-based protection is provided by many external security devices, but it is not always effective as a MIMO-based attack can recover jammed signal fully or partially~\cite{tippenhauer2013limitations}. 

\vspace{-0.3cm}    
\subsection{Data Protection Mechanisms for Communication and System-level Attacks}
    
The data protection mechanisms ensure the safety of important medical information from unintended corruption, compromise, or loss regardless of its physical or logical location. Though the most common solution for data protection is encryption, selecting an appropriate cryptographic algorithm for resource-constraint healthcare devices is important. In this regard, machine learning and blockchain-based approaches can be alternatives to encryption that can protect health data from tampering and detect attacks in healthcare systems. To defend against system-level attacks, access control mechanism is a good solution to protect healthcare data from unauthorized users.
    
\vspace{6pt}
\noindent\textbf{Encryption:} For transmitting sensitive data, especially over a wireless channel, encryption is a fundamental technique to secure the communication channel. Although there are many encryption techniques to follow, high energy consumption and implementation costs of encryption are big issues for resource-constraint medical devices~\cite{hosseini2011lightweight, rostami2013balancing}. 
As symmetric encryption mechanism usually consume less power than asymmetric encryption mechanism, it is more practical to use then for the resource-constrained platforms like medical devices~\cite{potlapally2003analyzing}. In this domain, \textit{PRESENT} and \textit{KATAN}, two lightweight hardware-oriented block ciphers \cite{bogdanov2007present, de2009katan} were developed with 64-bit block size and 80-bit key. Additionally, a low power block-cipher based security protocol was proposed in~\cite{beck2011block}, which offers two operational modes: stream and session mode. For short message transmission, the stream mode utilizes output feedback (OFB) to obtain a scalable stream cipher that enables strict duty cycling for energy. The session mode utilizes the CBC and a challenge-response scheme to provide advanced security to health data. 
 
Symmetric encryption algorithms depend on the distribution of shared keys between devices where healthcare devices often have to communicate with previously unknown devices. If the patient is unconscious, then key distribution needs to be done without the patient's intervention for timely treatment. The secret key can be imprinted on a card or on wearable devices where it can be hidden secretly \cite{denning2010patients} or printed directly onto the patients' skin using ultraviolet-ink micro-pigmentation that is only visible under ultraviolet light \cite{schechter2010security}. Even if the secret key is shared secretly or pre-loaded, it needs to be changed periodically, and new keys should be generated with a secure agreement protocol, high randomness, and minimum energy overhead. For instance, RSSI is the symmetrical property of the wireless channel between two devices that can generate a symmetrical key from the communication \cite{shi2013ask, ali2012zero, jana2009effectiveness, mathur2008radio}. These techniques are useful not only for the initial key sharing setup but also for renewing the key periodically to prevent eavesdropping.
   
However, even low power symmetric ciphers may increase power consumption, which will shorten the battery life. To address this issue, Different compression techniques can be used before applying encryption to reduce the power consumption and transmission cost \cite{zhang2014trustworthiness}. Compressive sensing is well suited since compression can be realized with very low computational and energy footprint~\cite{donoho2006compressed}. Zhang et al. presented \textit{Encompression}, a combination of compressive sensing, encryption, and integrity checking, which utilizes the sparsity of sensor data for reducing power consumption in medical devices~\cite{zhang2013energy}. Researchers showed that \textit{Encompression} can reduce 78\% compared to traditional encryption and integrity checking with a reasonable compression ratio of 6-10x. The AES \cite{heron2009advanced}, and the secure hash algorithm (SHA) \cite{bertoni2013keccak} are used for data integrity and confidentiality where a hash algorithm is used on original data. As a result, imposters cannot generate encrypted data without knowing the AES secret key. Bu et al. \cite{bu2019bulwark} presented a mac-then-encrypt (MtE) security mechanism combining with AEC-CBC mode to protect IMD from communication-based attacks. As current IMDs are equipped with AES using 128 or 192- bit encryption keys, the IMD's information (health data from sensors to controllers), the timestamp, and the authentication signature can be wrapped all under 128 bits or 192 bits depending on the demand. As a result, MtE adds no extra transmission overhead in IMDs equipped with 128 or 192-bit AES-CBC mode.

\vspace{6pt}
\noindent\textbf{Machine learning-based approaches:} Machine learning (ML) is a data analytic technique that provides healthcare systems the ability to learn from data and perform specific tasks such as anomaly detection, behavior analysis, etc. from experience without being explicitly programmed. ML algorithms have been explored widely by the research community to detect attacks on healthcare systems~\cite{rathore2020deep, sikder2019patent}. Saeedi et al. demonstrated how a decision tree algorithm can be used to detect malicious attacks in healthcare devices~\cite{saeedi2019machine}. Here, authors used the normal behavior of healthcare devices as ground truth, and any deviation from the normal behavior was identified as an attack. Vhaduri et al. used several physiological and behavioral parameters such as calorie burn, average step counts, minute heart rate as features of support vector machine (SVM) to detect unauthorized access to a healthcare device, and its captured data~\cite{vhaduri2017wearable}. In a recent work, Newaz et al. proposed \textit{HealthGuard}, a novel ML-based security framework to detect malicious activities in a connected healthcare system~\cite{newaz2019healthguard}. \textit{HealthGuard} collects the vital signs of different healthcare devices and uses several ML algorithms to correlate the changes in body functions of the patient to distinguish benign and malicious activities. Although ML algorithms can identify anomalous behavior in a healthcare system, implementing an ML-based solution on medical devices can consume more energy, which is an issue for these resource-constrained devices.

\vspace{6pt}
\noindent\textbf{Blockchain-based approaches:} A blockchain is a distributed system that maintains a continuously growing list of data records and keeps these records safe from tampering. Blockchain-based security frameworks are widely used by researchers to protect healthcare data from unauthorized entities. Chen et al. proposed a blockchain-based storage scheme for medical data to ensure safe data storage and sharing~\cite{chen2019blockchain}. Researchers also introduced a service framework for sharing medical records to describe the process of personal medical data management in some applications. A novel hybrid approach introduced in \cite{dwivedi2019decentralized} that combines several cryptographic primitives (e.g., private key, public key, blockchain, etc.) to develop a patient-centric access control for electronic medical records. Srivastava et al. introduced a reliable data communication between the network and storage with more advanced and lightweight cryptographic techniques like the ARX encryption scheme~\cite{srivastava2019light}. They introduced the concept of Ring Signatures in the communication, which provides important privacy properties like \textit{Signers Anonymity} and \textit{Signature Correctness}. The same group of researchers introduced \textit{GHOSTDAG}, a novel and unique blockchain protocol for remote patient monitoring, which uses a directed acyclic graph instead of classic long singular blockchains~\cite{srivastava2019data}. Researchers utilized the idea of smart contract programs from blockchain to monitor the health data of the patients. However, the ownership of the current medical data is still an issue, and currently, there are no rules for using blockchain in the health insurance portability and accountability act. 

\vspace{6pt}
\noindent\textbf{Access control mechanisms:} Access control mechanisms prevent unauthorized access to healthcare devices. Researchers proposed several types of access control for healthcare systems, including proximity-based, identity-based, role-based, attribute-based, and risk-based access control~\cite{sun2011hcpp, sikder2019multi}. 
Lin et al. proposed an identity-based access control to protect private health data in a cloud-assisted health monitoring system~\cite{lin2013cam}. In the role-based access control scheme, the service requester's role determines whether the access will be granted or denied. Li et al. proposed to give access rights to healthcare providers based on their roles in the wireless BAN~\cite{li2010data}. The attribute-based access control is an extension of identity-based access control where decisions are made based on a set of attributes (e.g., specialty, license validity, etc.). Attribute-based access policy has been employed in many research works to control access to medical data, BAN, cloud system \cite{li2012scalable, guan2015achieving}. The risk-based access control brings real-time, risk-aware decision-making capability in the access control mechanism. The anomaly detection-based access control schemes, fall into this category that constantly monitors any abnormal behavior in accessing a healthcare device \cite{hei2010defending}, \cite{zhang2013medmon}. Proximity-based access control is based on the distance between the programmer and the device \cite{halperin2008pacemakers, li2011hijacking}. Here, programmer can generate the same key to decrypt the communication only if it is near the patient. Fu et al. \cite{fu2019poks} presented a physical obfuscated key (POK)-based IMD access control mechanism where researchers leveraged IC cards of POKs for secure credential storage. They had designed a lightweight access control protocol with minimal computation and communication overhead on IMDs.

\vspace{-0.4cm}
\subsection{Limitations of Current Security Solutions}

In this subsection, we discuss the shortcoming of current security solutions, as well as highlight the limitations of healthcare devices that force the solutions to be too specific to be broadly useful. 
    
\begin{enumerate} [nosep, wide=0pt, leftmargin=*]

\item As third-party vendors manufacture most of the healthcare devices' hardware ICs, there is no specific standard to follow for IC manufacturing. As a consequence, an attacker can include an unproven intellectual property core into the IC that acts as an HT to perform malicious activities. As HTs introduces power and timing variation in healthcare devices. Existing hardware-centric solutions such as current analysis, delay variation, etc. are not suitable for HT detection.  

\item Currently, there is no base set of standards for the known behavior of healthcare device power consumption. Hence, the current malware detection techniques using power consumption analysis (e.g., \textit{WattsUpDoc} \cite{clark2013wattsupdoc}) are not an appropriate choice as it can raise false alarms and have a low accuracy rate in detecting malware. Also, most of the healthcare applications' source code is not open source, which makes it challenging to use static analysis for finding software-level bugs (e.g., buffer overrun, initialized variables, null pointer dereference, etc.) and defend against corresponding attacks.

\item The healthcare devices and the programmers mostly use a fixed secret key loaded during the manufacturing time. Using the same pre-shared key for an extended period increases the possibility of a successful cryptanalysis attack. To solve this problem, researchers proposed simultaneous LED blinking to be used as an authentication mechanism~\cite{li2013secure}. However, this technique is not suitable for emergency situations.
        
     
\item As most of the medical devices (e.g., implantable devices, wearable devices, etc.) are resource-limited, it is not easy to implement any cryptographic algorithm on the devices. One solution can be the use of cryptographic key computation from the patient's vital state~\cite{xu2011imdguard}. However, the computation of the key also consumes high power and reduces battery life.  

\item Medical devices collect information from multiple sensors (e.g., blood pressure, glucose, motion, etc.) to observe different vital signs of a patient. In real life, patients could become unconscious without showing any alarming symptoms, and the emergency personnel would require access to healthcare devices like IMDs to collect information related to previous health-related data. Researchers proposed to use a specific key that is shared between a common group of people like doctors and emergency personnel as a backdoor solution \cite{hei2010defending}. However, a privilege escalation attack can easily reveal the secret key, and attackers can obstruct the treatment plan causing a life-threatening situation.

\item Current software verification tools are written in a high-level programming language, which are not suitable for platform-specific and low-level applications of medical devices. These programs have to interact with medical sensors, actuators, and other hardware peripherals. Also, system-level properties need to be verified with real-world healthcare device interfaces \cite{li2013improving}.

\item Different healthcare devices are using different communications protocols, software, and application platforms for performing their healthcare-related tasks. Hence, it is hard for security researchers to provide common security solutions for healthcare systems. Although a recent study proposed a standard solution \textit{HealthGuard} \cite{newaz2019healthguard} to find anomalous activities in a healthcare system, the device dependency in this solution is too complex to consider in the current healthcare domain.
\end{enumerate}

\vspace{-0.4cm}

\section{Discussion and Future Research Directions}\label{discussion}

In this survey, we focus on the primary security and privacy goals for the next generation of healthcare devices and analyze the most common and related protection mechanisms proposed so far. To secure a healthcare system, security proposals must consider the energy, storage, and computing power constraints of the healthcare devices. Furthermore, the security solution must maintain the balance between patient safety and the security level offered. In this section, we discuss the security recommendations and practices, so the future research directions, that are required to be addressed for ensuring security and privacy in healthcare systems: 
    
    
    \noindent\textbf{Securing healthcare data:} The healthcare industry is generating data rapidly, and this healthcare data have clinical, financial, and operational value in the market. To protect this healthcare data from breaches, effective security measures like cryptographic solutions are needed for securing wireless communication, as well as the information stored in the device or the server. In this regard, cryptographic protocols that are symmetric \cite{malasri2009design} and lightweight \cite{hosseini2011lightweight} can provide a means to control access to healthcare devices and protect against spoofing and elevation of privilege attacks. However, incorporating cryptographic mechanisms in existing healthcare devices implies that current devices like IMDs must be replaced or redesigned. In the case of an emergency, communication with unauthorized personnel may be needed, which can be interrupted due to implemented cryptographic schemes. Hence, researchers should focus on developing medical-centric cryptographic solutions to meet the unique security goals of healthcare system. 

    \noindent\textbf{Lack of standard communication protocols:} Communication standard for healthcare devices is a good practice, 
    and many international standards are considered as prerequisites for the certification of healthcare devices. These standards are limited to the development and design risk assessment process, but not focused on the specific security requirements within the sophisticated deployment setting. Many security flaws and corresponding vulnerabilities like SQL injection and buffer overflow are a consequence of poor software design, which may be related to communication standards used in those devices \cite{khera2017think, sikder2018iot}. The design aspects of different medical standards, such as 62304/82304/80002, are crucial for cybersecurity, and are briefly described below:
    
    \begin{itemize} [nosep, wide=0pt, leftmargin=*]
        \item IEC 62304:2006 - Medical device Software -- Software life cycle processes define the life cycle requirements for medical device software and software used within the medical devices.  It establishes a common framework for the medical device software life cycle process. This standard is currently under revision and adjustment with ISO 82304.
        
        \item ISO/IEC 27032:2012 - Information technology -- Security techniques -- It provides cybersecurity guidelines to improve the state of security. 
        It brings out the other aspects of security like information security, network security, internet security, and critical information infrastructure protection, etc. to highlight the essential practices in cybersecurity.
        
        \item IEC 82304-1:2016  Health software -- Part 1: General requirements for product safety is a standard for the security of health software products designed to operate on general computing platforms (an evolution of IEC 62304). This standard works when health software is part of - or embedded in - a physical device \cite{williams2015cybersecurity}. 82304 and 62304 both focus on the process of product design, software validation, maintenance, and testing.

        \item ISO/IEC 8001 (Risk Management of Medical Devices on a Network) It defines the roles, responsibilities, and activities for IT-networks incorporating medical devices.
        
        \item IEC/TR 80002-1:2009  Medical device software -- Part 1: It provides the guidance for the application to comply with the requirements contained in ISO 14971:2007 and also provides direction for implementing a risk management process for medical device software, as part of the overall risk management process.  It is the principal standard for risk management regulation.
        
        \item ISO/TR 80002-2:2017 Medical device software -- Part 2: Validation of software for medical device quality systems is a technical report under development, which considers embedded and associated software with all medical devices. It includes many types of software used in device design, testing, component acceptance, etc.

        \item IEC/TR 80002-3:2014  Medical device software -- Part 3: Reference model of medical device software life cycle (IEC 62304) defines the software life cycle processes 
        and associated safety class definitions that derives from IEC 62304.
        
    \end{itemize}
    
    Although following the standards mentioned above is a good practice in development life-cycle processes, but they do not deal with the fundamental cybersecurity protection required for the medical devices in healthcare systems. Hence, the future research direction should enforce on selecting a common communication protocol standard so that researchers can provide a universal solution for any threats to healthcare communication medium and devices.

\noindent\textbf{Fault-tolerant design:} Reliability is the top priority in life-critical healthcare systems. During the time of manufacturing typically, it is possible to identify a large number of hardware or software defects, but exhaustive testing and complete fault coverage may not be possible. Through the simultaneous detection, diagnosis, and correction of fault effects, fault-tolerant designs enable a system to continue operating in the event of faults in its components. Additionally, it can be extended to cope with software errors caused by design inadequacies \cite{randell1975system}. Although some types of redundancy like time, hardware, or information are required for fault tolerance, this redundancy often costs performance degradation or other overhead. Trip modular redundancy (TPM) can be a good example, which employs three copies of a module and uses a majority vote to determine the final output \cite{lyons1962use}. Although it costs three times more than the original circuit, fault-tolerant design techniques may be warranted in safety-critical medical devices.

\noindent\textbf{Intrusion detection mechanism to detect attacks:} Ensuring the safety of patients is the key reason for the need of securing healthcare devices. An intrusion detection system (IDS) is usually needed to detect any types of attacks. An alert can be generated to notify patients or medical staffs if an adversary is threatening the healthcare system or device. An IDS would monitor incoming traffic coming from the external device to the healthcare devices based on some predefined rules  (e.g., the delay between two successive requests coming from an external device, the length of the message payload, etc.)\cite{chatzigiannakis2007decentralized}. The research community should give more focus on developing a standard IDS where the predefined rules can be referred to different communication protocol standards in healthcare devices. 

\noindent \textbf{Fine-grained access control:} Current access control mechanisms, such as attribute-based policy \cite{li2012scalable} and risk-based access control \cite{sun2011hcpp} mostly focused on preventing unauthorized access to healthcare devices. However, these healthcare data transmit through different parts of a healthcare system, where the integrity and confidentiality of the data are not ensured. As a solution to this problem, a standard access control policy should be imposed on different components of a healthcare system to identify any security violation in data access.

\noindent\textbf{Lack of general platform:} Different healthcare devices are using numerous hardware specifications, application software, communication protocols, and OSes. This heterogeneity makes it hard for security researchers to study existing threats and provide a common security solution for these healthcare devices. As a result, most of the security solutions are platform-specific and unable to solve the wide-range of security problems. Researchers, developers, and industry should work on developing a common standard for healthcare systems that will aid researchers in developing generalized security solutions.

\noindent \textbf{Privacy-preserving healthcare system:} Existing privacy solutions such as user and communication anonymity \cite{dwivedi2019decentralized, srivastava2019light, srivastava2019data} mostly focused on preventing any unauthorized access to the healthcare data and communication channel. However, there are still several issues in healthcare systems that could affect the privacy of the users and healthcare data if not addressed properly. For instance, several healthcare devices have physical addresses written in their body which violates the device anonymity requirement for the healthcare system. To ensure privacy, manufacturers should find a safer way to share device information with users, and researchers should put more focus on imposing a standard privacy policy on different components of the healthcare systems.

\noindent\textbf{Study of smart medical devices and existing threats:} In recent years, the concept of IoT has been integrated into the medical domain, making modern medical devices context-aware and smart. However, there are several attacks on IoT devices that can be adapted in a smart medical platform with minimal modifications. As the security requirements of IoT devices and healthcare systems vary a lot, existing security solutions of IoT devices often can not ensure the end-to-end security need of medical IoT devices. Hence, it is necessary to study the smart healthcare platforms, as well as the medical IoT domain to identify the underlying threats and develop security solutions specific to the healthcare domain. The research community should give more focus on smart healthcare devices to address these threats, and industry experts should propose a standard practice for device manufacturing and application development to minimize these threats.

\vspace{-0.2cm}

\section{Conclusion}\label{conclusion}
\vspace{-0.2cm}

In this paper, we presented an overview of the existing security and privacy research in healthcare systems. 
Increasing functional complexity, more software programmability, and growing wireless network connectivity are general trends observed in healthcare devices applications. However, there are side effects of these trends, which makes the healthcare devices and applications increasingly vulnerable to security and privacy issues. 
We analyzed various aspects of the threats and how the current solutions 
overcome these threats. Given the critical tasks performed by healthcare devices, these issues should be addressed aggressively and proactively by the community. 
We believe this survey will have a positive impact on the 
medical community by documenting recent attacks and defenses 
and facilitating 
a more aware ecosystem for the security and privacy in healthcare systems.


\section{Acknowledgment}\label{sec:acknowledgment}
This work is partially supported by the US National Science Foundation (Awards: NSF-CAREER-CNS-1453647, NSF-CNS-1718116) and Florida Center for Cybersecurity's Capacity Building Program. The views expressed are those of the authors only, not of the funding agencies.
\vspace{-2pt}

\bibliographystyle{IEEEtran}
\bibliography{acmart}

\end{document}